\documentclass[]{sig-alternate-10pt}

\usepackage[numbers,sort&compress]{natbib}

\usepackage{amssymb}
\usepackage{amsmath}
\usepackage{subfigure}
\usepackage{epstopdf}
\usepackage{multirow}
\graphicspath{{Figs/}}

\begin{document}
\title{Isolating Mice and Elephant in Data Centers}
\author{Wenxue Cheng$^\star$, Fengyuan Ren$^\star$,  Wanchun Jiang$^\dag$, Kun Qian$^\star$, Tong Zhang$^\star$, Ran Shu$^\star$\\
$^\star$Dept. of Computer Science and Technology, Tsinghua University, Beijing, China\\
$^\dag$Dept. of Computer Science and Technology, Central South University, Changsha, China\\
$^\star$\{chengwx, renfy, qiankun, zhangt, shuran\}@csnet1.cs.tsinghua.edu.cn, 
$^\dag$jiangwc@csu.edu.cn
}
\maketitle

\begin{abstract}
Datacenters traffic is composed by numerous latency-sensitive ``mice'' flows, which is consisted of only several packets, and a few throughput-sensitive ``elephant'' flows, which occupy more than 80\% of overall load. Generally, the  short-lived ``mice'' flows induce transient congestion and the long-lived ``elephant'' flows cause persistent congestion. The network congestion is a major performance inhibitor. Conventionally, the hop-by-hop and end-to-end flow control mechanisms are employed to relief transient and persistent congestion, respectively. However, in face of the mixture of elephants and mice, we find the hybrid congestion control scheme including hop-by-hop and end-to-end flow control mechanisms suffers from serious performance impairments. As a step further, our in-depth analysis reveals that the hybrid scheme performs poor at latency of mice and throughput of elephant. Motivated by this understanding, we argue for isolating mice and elephants in different queues, such that the hop-by-hop and end-to-end flow control mechanisms are independently imposed to short-lived and long-lived flows, respectively. Our solution is readily-deployable and compatible with current commodity network devices and can leverage various congestion control mechanisms. Extensive simulations show that our proposal of isolation can simultaneously improve the latency of mice by at least 30\% and the link utilization to almost 100\%.
\end{abstract}


\section{Introduction}
Experimental observations verify that the traffic in data centers follows ubiquitous heavy-tailed distribution: the majority of traffic is actually transferred by only a small number of long-lived flows (elephants), while the numerous short-lived flows (mice) only carry a minority of traffic. Most of flows are generally small in size (<10KB) and last a few to hundreds of milliseconds \cite{kandula2009nature}. The similar properties have also been observed by subsequent measurements \cite{benson2010network, greenberg2009vl2} and \cite{gill2011understanding}. Definitely, the traffic statistical characteristic is dominated by the services hosted in data centers. The real-time analytics \cite{engle2012shark, melnik2011dremel} and Online Data-Intensive (OLDI) applications like web search and online retail \cite{jalaparti2013speeding, decandia2007dynamo}, generate short messages, which is often bursty and latency-sensitive, whereas background traffic from data update, migration and backup may compete for network bandwidth in minutes or hours or more where throughput is far more important than latency. 

Because reducing application latency can improve user experience, and hence lead to significant financial loss revenue \cite{latency}, most of work particularly emphasizes on minimizing flow completion time or meeting deadline for short flows. The flow control as a fundamental congestion control mechanism does not receive enough attention. It has been shown that majority of congestion events are transient, and a small but significant fraction of core links appear to be persistently congested in data centers \cite{kandula2009nature, benson2010network}. Generally, the transient congestion is caused by short-lived flows, and the persistent congestion is due to long-lived flows. Relieving congestion is preliminary for leveraging other traffic management schemes to satisfy special performance requirements.

Conventional wisdom tells that transient congestion can be coped with by the hop-by-hop flow control mechanisms properly, such as backpressure and pause, while the end-to-end flow control, such as TCP, is suitable to alleviate persistent congestion. Most of exited work is prone to employ the end-to-end flow control schemes relieve transient and persistent congestions together except that PDQ \cite{hong2012finishing} uses the pause mechanism to halt contending flows. The end-to-end flow control is reactive, and takes effect only after congestions occur. It typically requires a few RTTs to detect congestion and throttle source injection rate, by which time the congestion events is likely already over. However, the majority of congestion is transient caused by short-lived flows \cite{kandula2009nature, benson2010network}, which is responsible for over 90\% of packet loss \cite{kandula2009nature}. The resulting retransmissions further exacerbate an already congested network, and even impose heavy penalty on latency-sensitive short flows due to timeouts \cite{alizadeh2011data}. The hop-by-hop flow control can solve the problem of packet loss in the case of transient congestion, moreover can reduce short flow completion time since it can grad more bandwidth through aggressively sending packets at full port rate. Its drawback is the problem of head-of-line blocking resulting in the congestion tree. When congestion persists, the congestion tree quickly grows and precludes transmission of other flows that are not even destined for the congested area. While for transient congestion caused by short bursts, a congestion tree does not have time to propagate.

Although conventional wisdom reminds us that the hop-by-hop and end-to-end flow control mechanisms can effectively handle transient and persistent congestions, respectively, are they beneficial, neutral or detrimental in the context of long and short flows traversing in same queue, particularly for typical mixed traffic comprising both mice and elephants with obvious statistical characteristic in large-scale data centers? To the best of our knowledge, there are few investigations on this fundamental issue. Motived by this concern, in this work, we will provide an in-depth understanding on how the hop-by-hop flow control interacts with the end-to-end flow control when mice and elephants meet in the same queue, and how this interaction affects the network performance and application's SLA, including latency and losslessness of mice and throughput of elephants. Inspired by these insights, we also suggest a readily-deployable solution to reconcile transient and persistent congestion caused by mice and elephants in data center networks. The main contributions are two-fold:

(1)	Conducting experimental observations and analysis in typical scenarios where flow mechanisms and mixture traffic is configured in possible combination modes, we conclude that the hop-by-hop and end-to-end flow control mechanisms cannot harmoniously work together when numerous short-lived and a few long-lived flows are multiplexed in one queue. 

(2)	Our analysis enlightens a concise and straightforward solution on relieving transient and persistent congestion through isolating mice and elephants in individual queues and imposing hop-by-hop flow control and end-to-end flow control, respectively. This isolation has some additional advantages, including alleviating the spread of congestion tree inherent in hop-by-hop flow control scheme naturally, mitigating the bursty noise of congestion status in end-to-end flow control scheme,  endowing potential on reducing queueing latency for short-lived flows and improving throughput for long-lived flows. Last but not least, the solution is readily-deployable and compatible with the industry standards without any hardware changes.

The rest of the paper is organized as follows. In Section \ref{background}, we introduce the traffic patterns and their requirements in data centers as well as the DCB architecture in detail. And we  illustrate the performance impairment in the case of mixed mice and elephant using a simple simulation in Section \ref{motivation}. Section \ref{understand} studies the behaviors of PFC and QCN under different traffic patterns, and elaborately explains the performance impairment with mixed traffic patterns. Based on the implications in Section \ref{understand}, we propose a straightforward solution of isolating mice and elephant in individual queues and respectively imposing hop-by-hop and end-to-end flow control mechanisms in Section \ref{solution}. 
Section \ref{evaluation} evaluates the proposed solution with both experiments and simulations. Finally, the paper is concluded in Section \ref{conclusion}.

\section{Background}
\label{background}

We firstly describe the traffic pattern and requirements in data centers, and then briefly introduce the Data Center Bridging (DCB) architecture.

\subsection{Traffic Pattern and Requirements}

\subsubsection{Mice and Elephant}
The traffics of many data center applications such as web services and key-value stores are mainly made up of short messages, which are small in size (<10KB) and accordingly called mice flows. Moreover, the mice flows always appear in the burst pattern.
For example, measurement results in \cite{atikoglu2012workload} show that the traffics in Memcached systems exhibit obvious shortness and burstiness.
In addition, these bursty mice flows always show the large-scale property, not only in the sense of numerous covered physical nodes, but also in the high concurrency.
Take the Memcached clusters of Facebook for example, they hold tens of thousands of servers and more than $100K$ requests are activated per second \cite{nishtala2013scaling}.
However, large-scale as these flows are, they only occupy a small fraction of network resources in total. According to prior researches \cite{alizadeh2011data, benson2010network, gill2011understanding}, the average load of mice flows is no more than 5\%. It is the long-lived elephant flows, which are generated from operations such as backup, update, or VM migration, that grasps more than 80\% of bandwidth in data centers.
In brief, data center networks are in the face of the mixture of elephant and mice.


\subsubsection{Requirements}
Because reducing application latency can improve user experience, leading to significant financial revenue \cite{latency}, flow completion time (FCT) or deadline of mice flows is the most concerned performance metrics.
Recently, FCT has been improved with various methods,
such as employing dedicated optical or wireless links to offload elephant flows \cite{wang2011c, farrington2011helios, halperin2011augmenting, hamedazimi2014firefly}, load balancing \cite{al2010hedera, benson2011microte, alizadeh2014conga, he2015presto}, differentiated services based on priority \cite{hong2012finishing, alizadeh2013pfabric, grosvenor2015queues, bai2015information} and customizing transport protocol \cite{alizadeh2011data, vamanan2012deadline, wilson2011better, mittal2015timely, zhu2015congestion}.
And the latency of each single packet in mice flows is emphasized, especially with the popularity of key-value store and real-time streaming computing, where mice flows contain only one or two packets.
Generally, the latency of each packet is composed of the processing time in the network stack, the queueing time in NICs and switches, as well as the transmission and propagation time on hardware devices. Specifically, when packet loss occurs, the retransmission time is also included.
Because the retransmission time can be as large as the sum of all other parts, the packet loss becomes intolerable.
Moreover, technologies supposed to be deployed over lossless fabrics, such as Remote Direct Memory Access (RDMA), is widely adopted to reduce the CPU consumption and the processing time in network stacks. In this context, mechanisms preventing packet losses are strongly needed in current data centers.


Different from mice flows, in view of elephant flows the throughput is far more important than latency, because an elephant flow may compete for network bandwidth in minutes or hours or more. Therefore, it's crucial to avoid link underutilization for elephant flows.
\begin{figure}[t]
	\centering
	\includegraphics[width=1\linewidth]{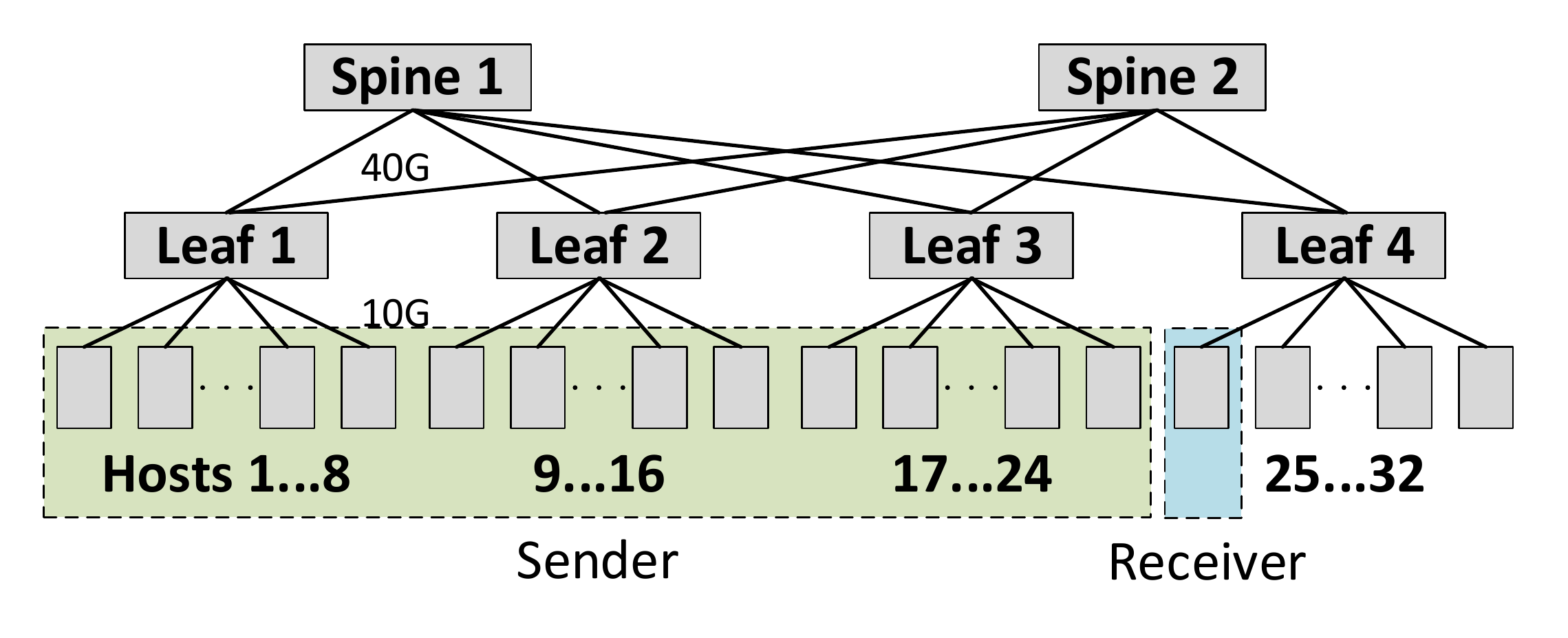}
	\caption{Many-to-one scenario.}
	\label{fig:topology}
\end{figure}

\subsection{Data Center Bridging Architecture}
Traffic management is crucial in satisfying losslessness and low latency of mice flows and high throughput of elephant flows.
Currently, the IEEE 802.1 Data Center Bridging (DCB) task group has developed PFC to prevent frame losses as well as QCN to eliminate the saturation tree problem generated by PFC for Ethernet \cite{ieee8021dcb}.
Detailed enhancements of Ethernet defined by DCB are as follows.
\subsubsection{Hop-to-hop flow control (PFC)}
The Priority-based Flow Control \cite{ieee8021qbb} is developed to enable lossless semantics required by some protocols in SAN and HPC, such as FC (Fibre Channel) and IPC (Interprocess Communication). PFC devises a hop-by-hop flow control mechanism to proceed/halt packets in Xon/Xoff pattern with a set of priorities predefined according to traffic class, for example HPC, SAN and LAN.
That is, once the ingress queue grows exceed a stated threshold $K1$, a PAUSE message is sent to the upstream device and the congested switch will send a RESUME message when the queue falls bellow a lower threshold $K2$.
To avoid packet drops before PAUSE message arrive and take effect on the upstream device, the switch should reserver enough buffer exceed the threshold $K1$.
The main problem of PFC is the congestion tree that precludes transmission of other flows that are not even destined for the congested area.
\subsubsection{End-to-end flow control (QCN)}
The Quantized Congestion Notification \cite{ieee8021qau} scheme is defined to manage long-lived flows within network domains supporting services that require limited bandwidth delay. Specially, QCN can suppress the spreading of congestion tree caused by persistent congestion. In reality, compromising between the cost and complexity of QCN implementation and its necessity in small or medium-scale private data centers, some product vendors decide to temporarily give up to support QCN \cite{what}.

It is worth notice that the QCN is an sampling-based congestion control mechanism. Once the feedback value $F_b\triangleq Q_{off}+\omega Q_\delta$ is positive, the congestion point (CP) will send the congestion notification message (CNM) to the source. $Q_{off}$ is the difference between actual queue length and the expected value and $Q_\delta$ is the difference between two successive sampled queue length.

Thus the DCB architecture is a typical scenario where the hop-to-hop and end-to-end flow control mechanisms work together. We will study on the performance of mixture traffics under DCB architecture.

\section{Performance Impairment}
\label{motivation}

\begin{figure}
	\centering
	\subfigure[Latency of mice flows]{
		\includegraphics[width=0.45\linewidth]{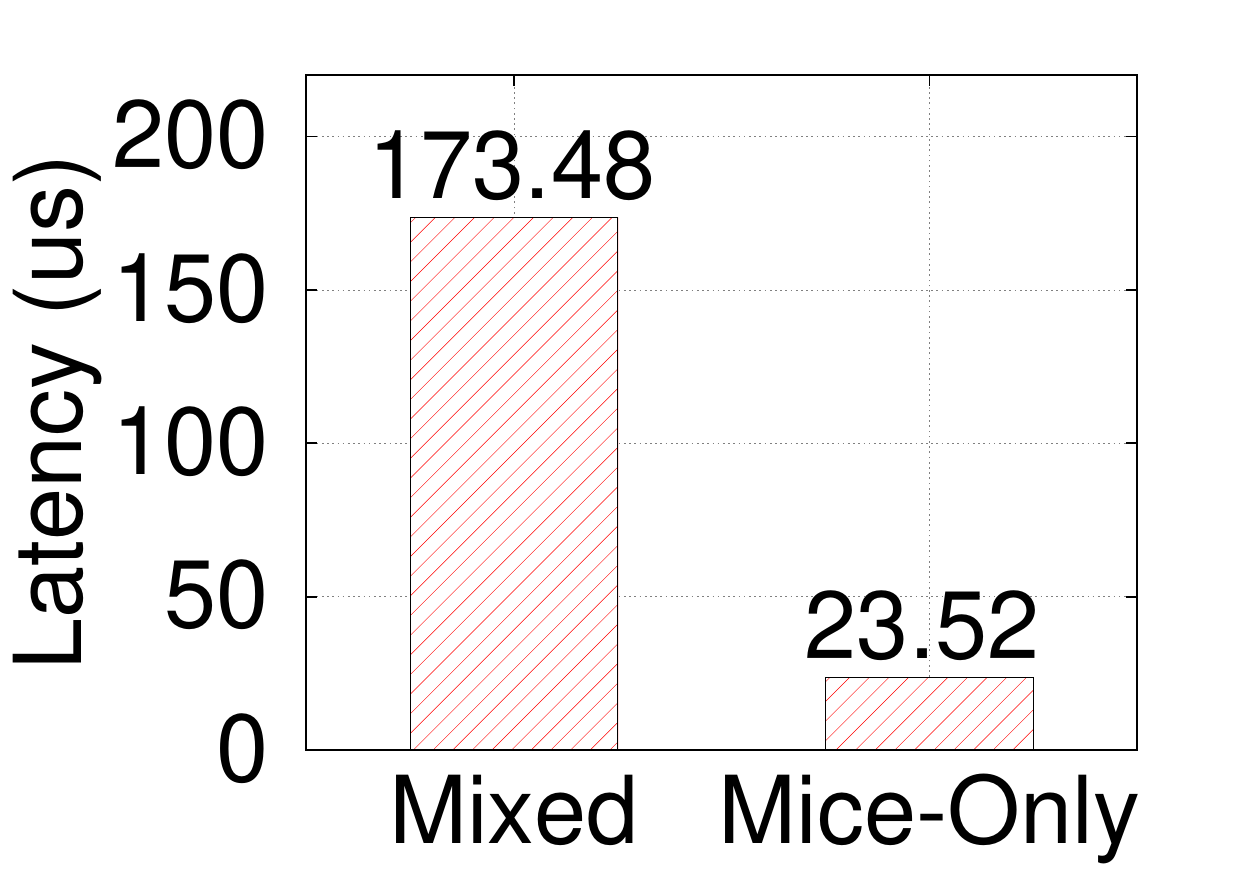}
		\label{fig:m-latency}
	}
	\hfill
	\subfigure[Bottleneck throughput]{
		\includegraphics[width=0.45\linewidth]{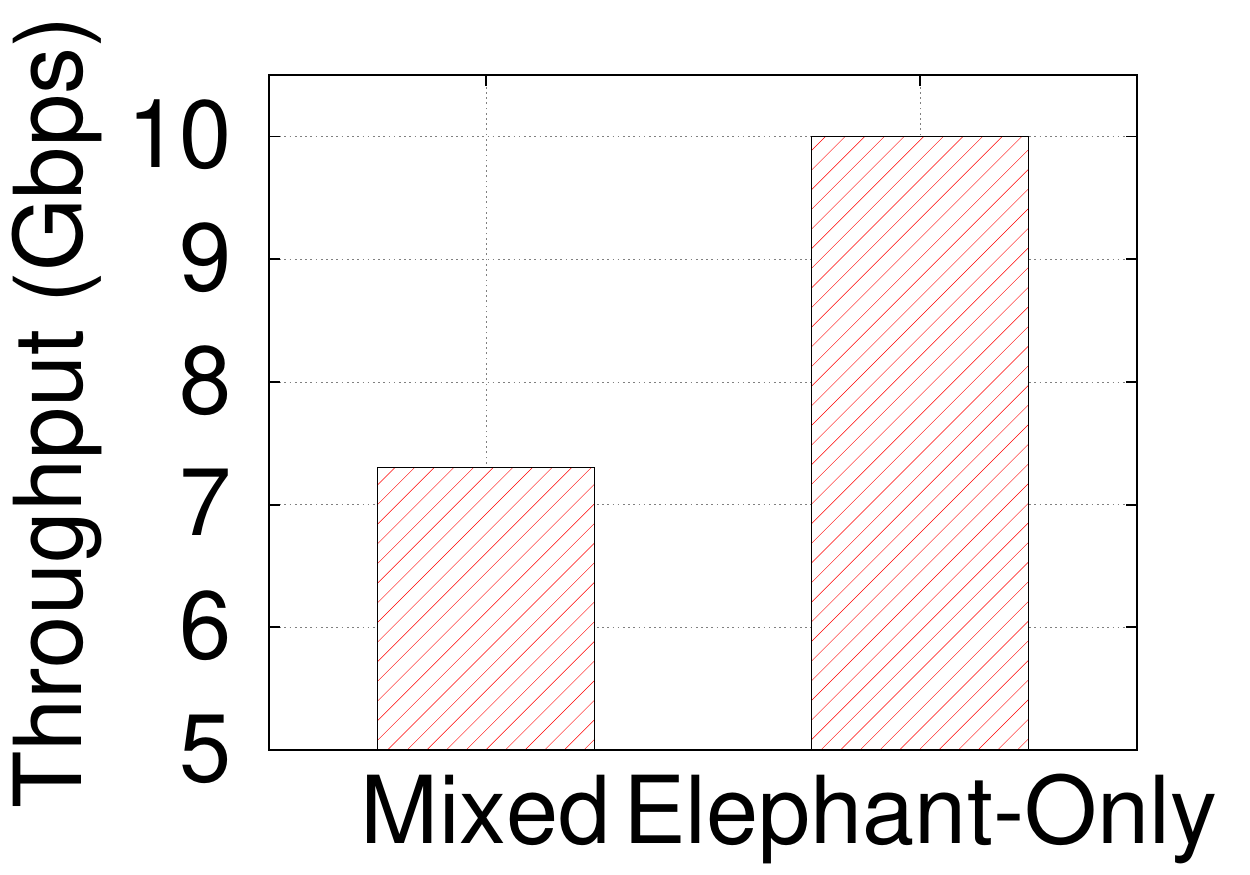}
		\label{fig:m-utilization}
	}
	\caption{Combination of PFC and QCN}
\label{fig:impairment}
\end{figure}

To examine the performance of DCB with the mixture of mice and elephant flows in datacenters, we conduct a simple simulation with the simple leaf-spine topology illustrated in \emph{Fig.}\ref{fig:topology}.
In simulation, the traffic follows a many-to-one pattern, i.e. all senders simultaneously transmit data to the receiver.
In view of each leaf switch, the traffic consists of mice and elephant flows, and the load of mice flows accounts for no more than 5\%. Both PFC and QCN are enabled.

We measure the latency of mice flows, utilization of the bottleneck link, and the queue length in the rightmost Leaf switch.
For comparison, we also carry out the simulation but using only mice or elephant flows in the same topological circumstance.
As shown in \emph{Fig.}\ref{fig:impairment}, in the mixed situation, the latency of mice flows is much larger than that in the mice-only scenario in average, while the bottleneck throughput smaller than that in the elephant-only scenario.
Although DCB try to provide a zero-loss and congested-less fabric, it fails to simultaneously provide satisfactory small latency for mice flows and high throughput for elephant flows in the mixed scenario. As analyzed in the next section, this performance impairment is induced by the misunderstanding on the impacts of the mixture of mice and elephant.


\section{Mixture of   Mice and Elephant}
\label{understand}
To explain above performance impairment with the mixture of elephant and mice, we study the behaviors of PFC and QCN under different traffic patterns, and show that the interaction of elephant and mice should be treated carefully regardless of the traffic management schemes.

\subsection{Hop-by-hop flow control (PFC)}
\begin{figure}[t]
	\centering
	\subfigure[Latency of mice flows]{
		\includegraphics[width=0.45\linewidth]{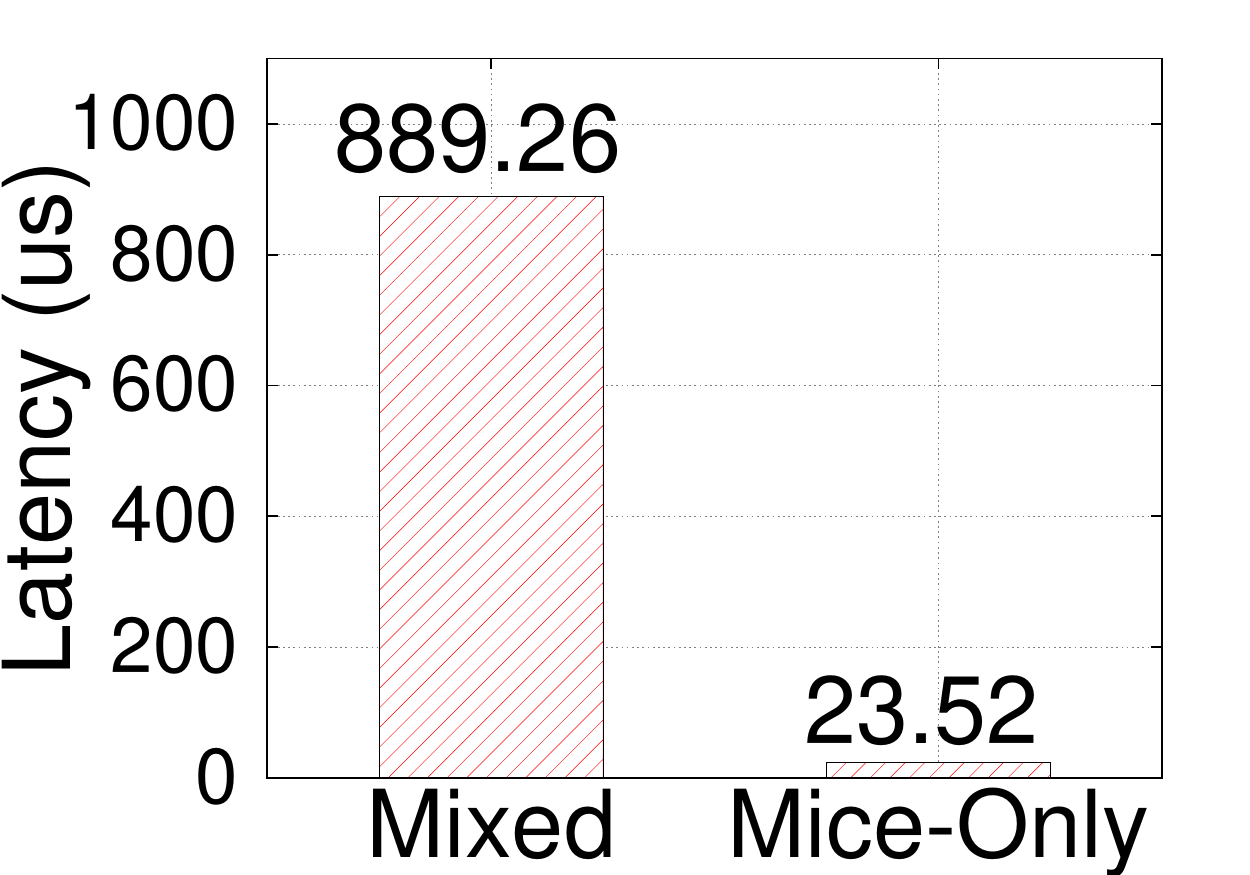}
		\label{fig:pfc1-latency}
	}
	\hfill
	\subfigure[Bottleneck throughput]{
		\includegraphics[width=0.45\linewidth]{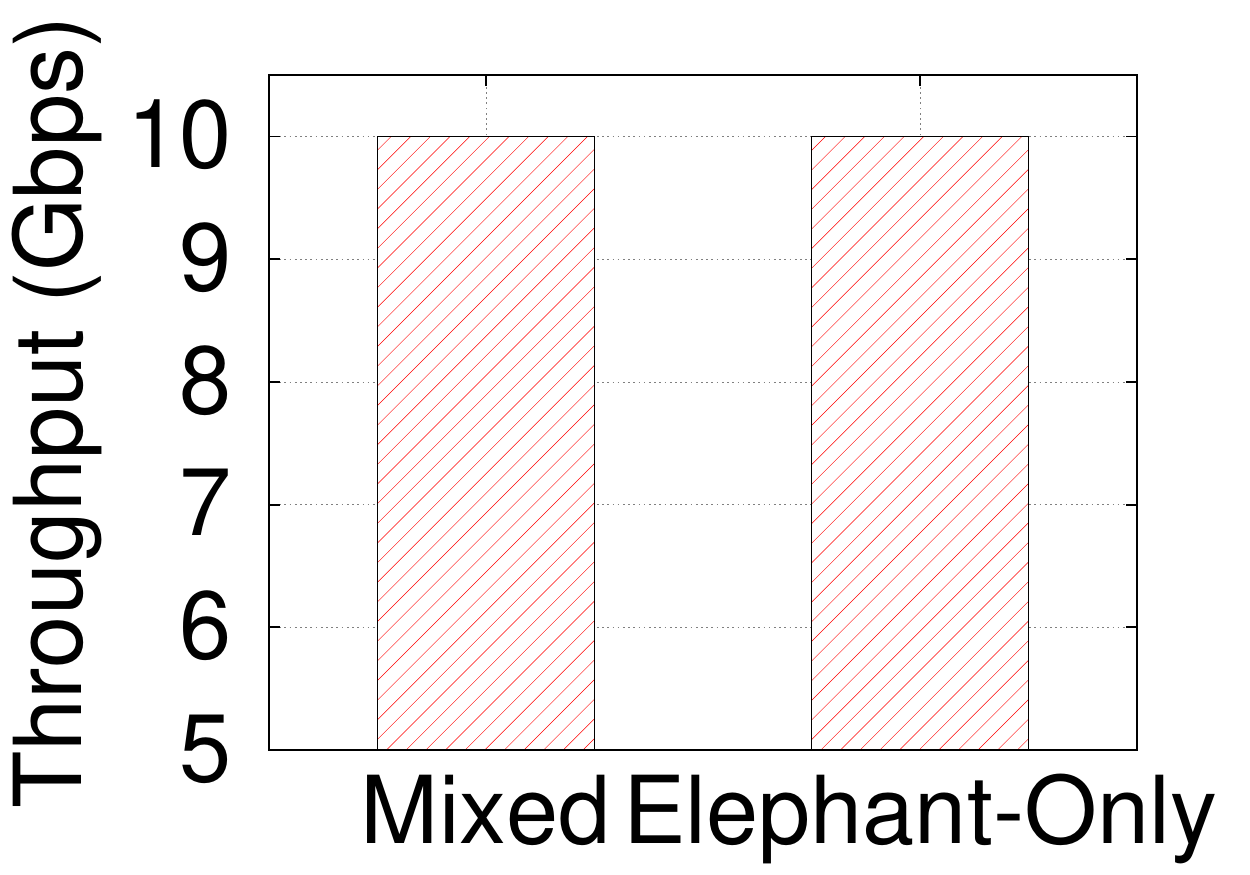}
		\label{fig:pfc1-throughput}
	}
	\subfigure[Queue Length]{
		\includegraphics[width=1\linewidth]{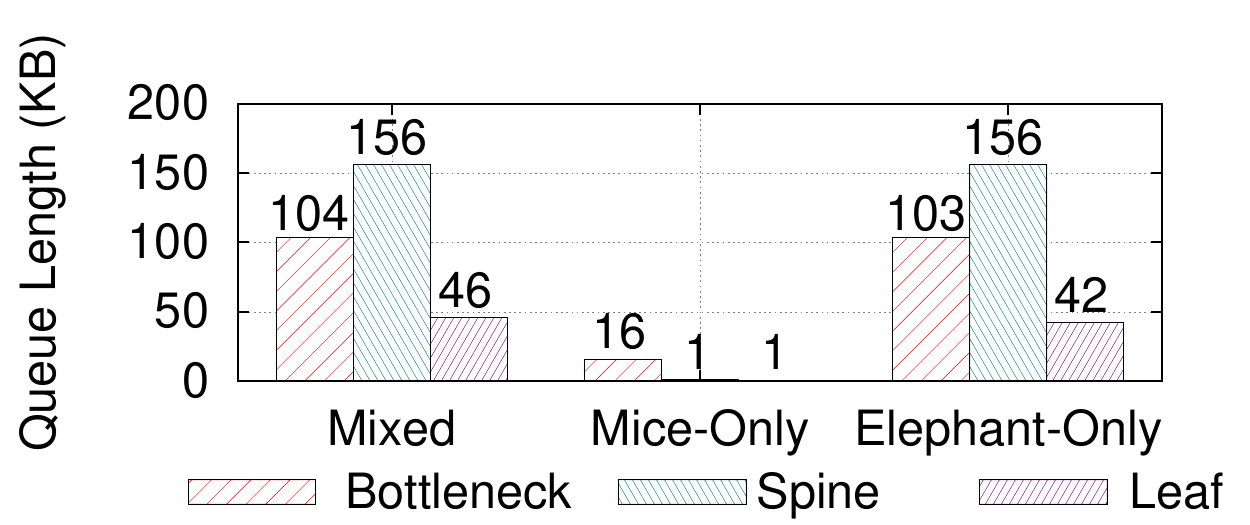}
		\label{fig:pfc1-queue}
	}
	\subfigure[Number of PAUSE events]{
		\includegraphics[width=1\linewidth]{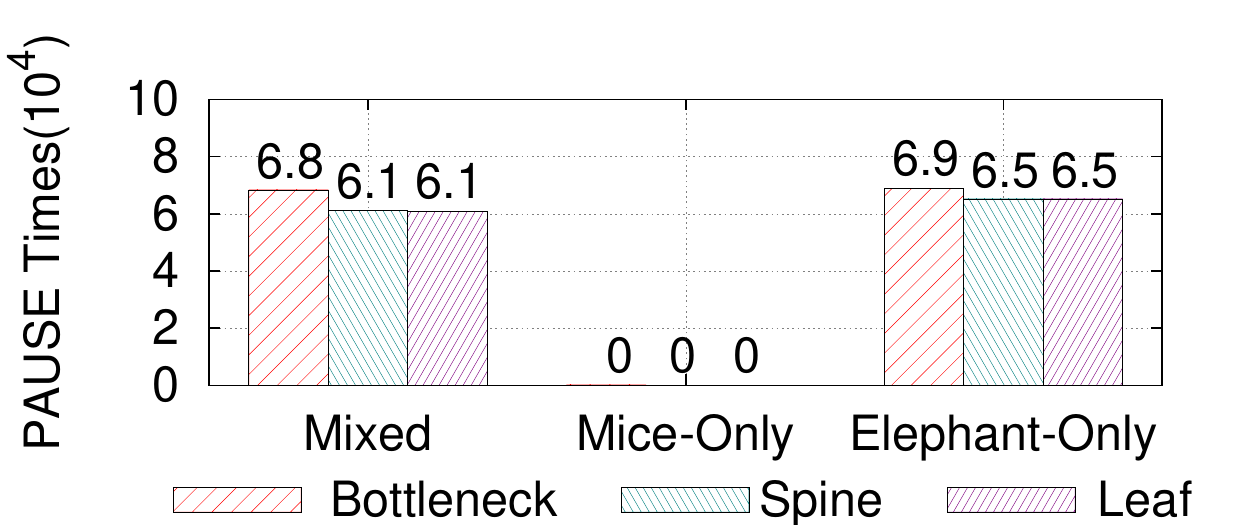}
		\label{fig:pfc1-pause}
	}
	\caption{PFC with many-to-one Traffic}
	\label{fig:pfc1}
\end{figure}

To study the impacts of PFC in above performance impairment, we firstly repeat the simulation in \S\ref{motivation} by enable only PFC, and then analyze it theoretically. The threshold of triggering PFC is $K=48KB$ for each ingress port.

\subsubsection{Simulations}
With only PFC, the simulation results are shown in \emph{Fig.}\ref{fig:pfc1}. Compared with the mixture scenario, although bottleneck link throughput is good enough in the elephant-only scenario, the latency of mice is greatly reduced in the mice-only scenario, as illustrated in \emph{Fig.}\ref{fig:pfc1-latency} and \emph{Fig.}\ref{fig:pfc1-throughput}.
To understand these phenomena, we measure the queue length in each switch.
Note that each switch has several ingress ports and the sum of queue length in all ingress ports is focused.
Moreover, as the topology is symmetric, the queue length of leaf switches are similar to each other, and thus we show the average value instead of each of them. So does the spine switches.
Obviously, $Leaf4$ is the only bottleneck switch in the many-to-one scenario. Therefore, the queue length of all other switches are expected to be $0$, excepting the bottleneck switch.
But the queue builds at all switches in the mixed scenario or the elephant-only scenario as presented in \emph{Fig.}\ref{fig:pfc1-queue}.
It indicates the PFC mechanism is triggered and the congestion is pressured back to spine switches and other leaf switches. In other word, the congestion tree is generated by PFC in the mixed or elephant-only scenarios.
The number of pause events summarized in \emph{Fig.}\ref{fig:pfc1-pause} confirms this indication.
This is the congestion-spreading issue of PFC.
In contrast, in the mice-only scenario, there are only a small queue at the bottleneck switch and no pause events, as shown in \emph{Fig.}\ref{fig:pfc1-queue} and \emph{Fig.}\ref{fig:pfc1-pause}.

\begin{figure}[t]
	\centering
	\includegraphics[width=1\linewidth]{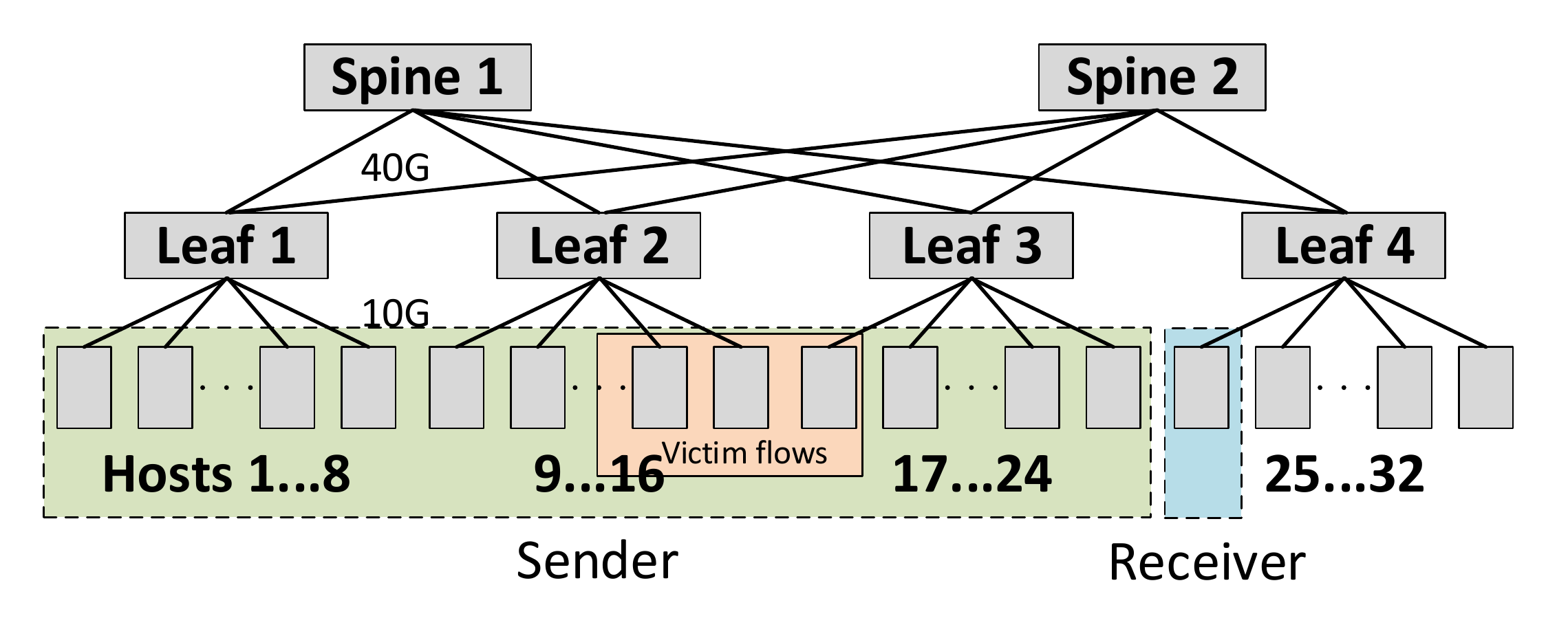}
	\caption{Head-of-line scenario.}
	\label{fig:topology2}
\end{figure}

Based on previous simulation, we add additional flows from $Leaf2$ to $Leaf3$, as shown in \emph{Fig.}\ref{fig:topology2}.
In this condition, links from $Leaf2$ to $Spine1,2$ become another two bottlenecks.
Repeat previous simulations, the results are shown in \emph{Fig.}\ref{fig:pfc2}.
Because the number of flows towards $Leaf4$ is originally larger than those towards $Leaf3$, it is normal that there exists a difference of about 6$\mu s$ in the latency of mice.
Similarly, in the mixed scenario the mice completion time almost doubles.
While the latency of mice remains favorable, in the elephant-only scenario, the utilization of bottleneck link $Leaf2$ to $Spine1,2$ is only about 40\%, as shown in \emph{Fig.}\ref{fig:pfc2-throughput}.
This is because PFC takes effect and frequently pauses all hosts under $Leaf2$.
In this way, the victim flows only get about $1Gbps$ bandwidth and the remaining $9Gbps$ is wasted.
Note that the victim flows have nothing to do with the congestion at $Leaf4$, but their rates are still limited by PFC so that the link bandwidth cannot be fully utilized.
This is fairness issue of PFC, generated by the congestion-spreading issue.

In sum, the harms of congestion-spreading and fairness issue of PFC are mainly on elephant flows. Actually, PFC does little harm to mice flows, as indicated in the mice-only scenario. This is because PFC is rarely triggered in the mice-only scenario, when the traffic load is not very heavy. Even if PFC takes effect, the paused hops are few in number, without hurting too many innocent flows in the mice-only scenario. The following theoretical analyses further verify these results.

\subsubsection{Analyses}

Let $K$ denote the PFC threshold and $Q$ be the queue length in the ingress port of the congested switch. Obviously, PFC is triggered when $Q>K$. During the time of transmitting the pause frame to the previous hop, there are still data arrival at the ingress port. Let $K_0$ denote the size of this part of data, $C$ denote the link capacity, and assume flows arrive according to a Poisson process with the average rate $\lambda$. Let $S$ be the random flow sizes, which obey the exponential distribution, then the service time of a flow could be expressed as $\frac{S}{C}$.
Define the average service rate as $\mu=\frac{C}{\mathbb{E}[S]}$, then the average link utilization is $\rho=\frac{\lambda}{\mu}$. These compose a M/M/1 queue model at the congested switch.

Let $w$ represent the average waiting time of packets in queue. Solving the M/M/1 model, we can get
\begin{equation}
\label{PFC}
\begin{array}{cl}
P\{Q>K\}&=P\left\{w>\frac{K}{C}\right\} \\[5pt]
&=e^{-\frac{C}{S}\times\frac{K}{C}(1-\rho)}=e^{-\frac{K}{S}(1-\rho)}
\end{array}
\end{equation}
This is the probability of triggering PFC in an ingress port.
As a step further, assume there are $n$ ingress ports at the congested switch in the many-to-one scenario.

The sufficient condition of triggering PFC simultaneously at $j$ ingress ports is
$Q>n^{j-1}K+\sum \limits_{i=1}^{j-1}n^{i-1}(K+K_0)$, and the corresponding steady state probability is
\begin{equation}
\label{PFC1}
\begin{array}{cl}
	& P\left\{Q>n^{j-1}K+\sum_{i=1}^{j-1}n^{i-1}(K+K_0)\right\} \\[5pt]
	=& P\left\{w>\frac{Q>n^{j-1}K+\sum_{i=1}^{j-1}n^{i-1}(K+K_0)}{C}\right\} \\[5pt]
	=& e^{-\frac{n^{j-1}K+\sum_{i=1}^{j-1}n^{i-1}(K+K_0)}{S}(1-\rho)}
\end{array}
\end{equation}
\begin{figure}[t]
	\centering
	\subfigure[Latency of mice flows]{
		\includegraphics[width=0.45\linewidth]{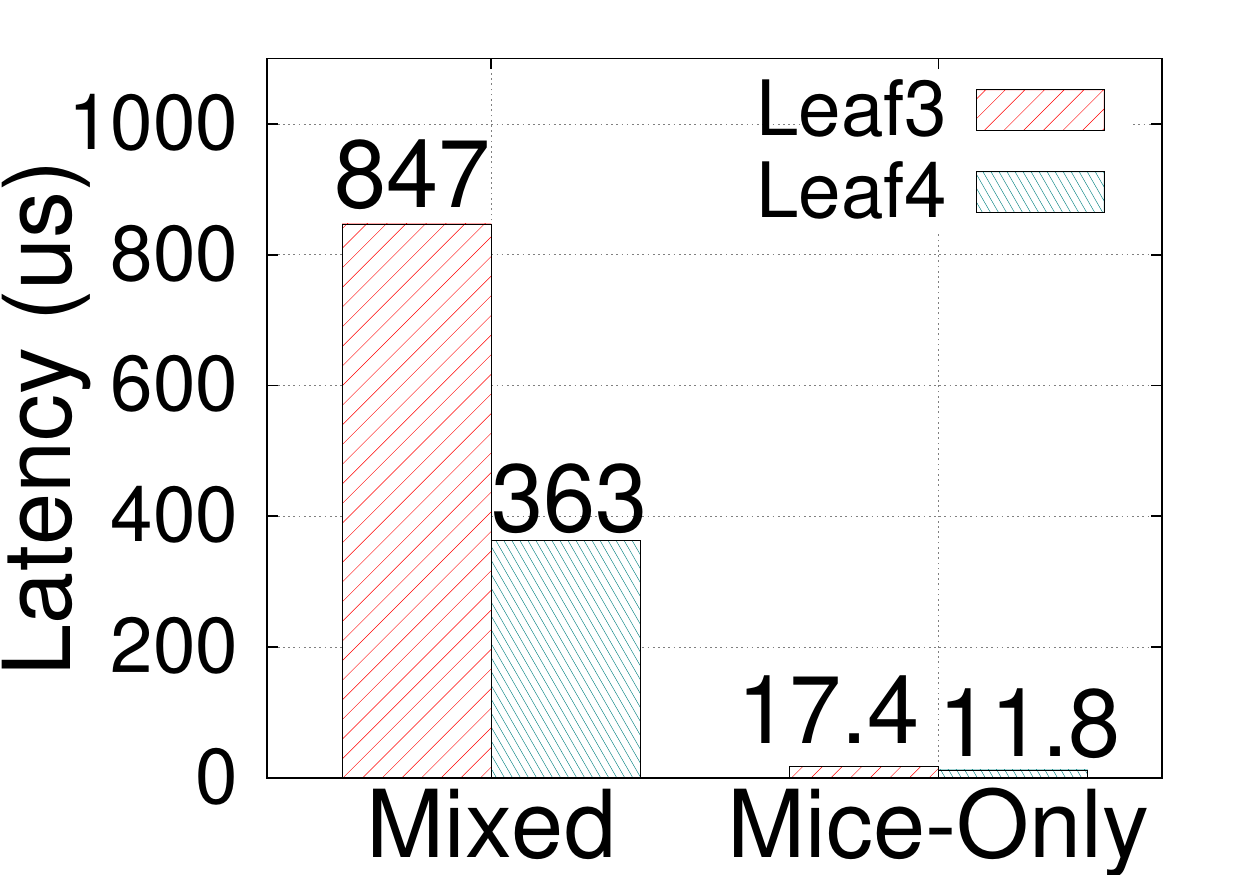}
		\label{fig:pfc2-latency}
	}
	\hfill
	\subfigure[Bottleneck throughput]{
		\includegraphics[width=0.45\linewidth]{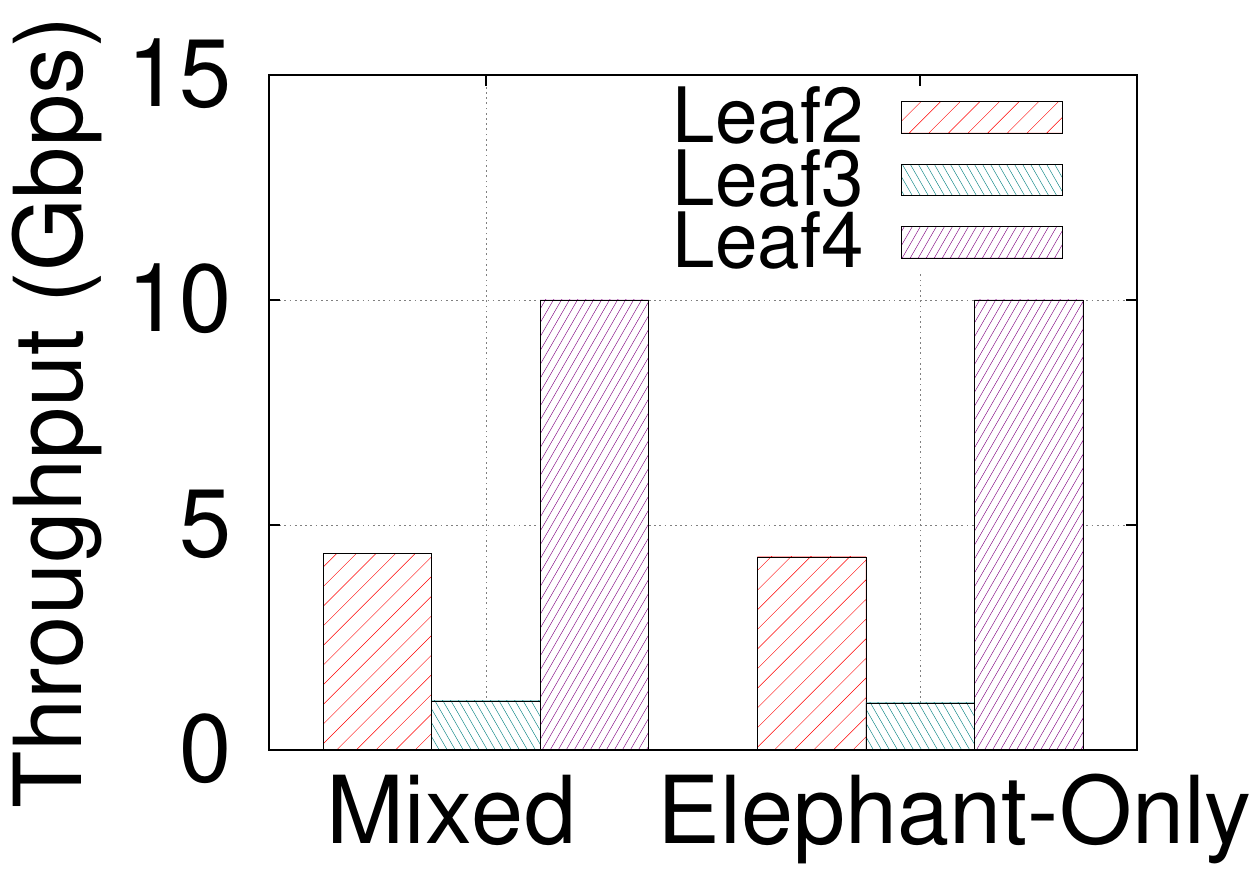}
		\label{fig:pfc2-throughput}
	}
	\caption{Fairness Issue of PFC}
	\label{fig:pfc2}
\end{figure}
Equations (\ref{PFC}) and (\ref{PFC1}) show that large buffers (large $K$), short flows (small $S$), and light workload (small $\rho$) all lead to fewer pauses. In practice, $\rho$ is smaller than 0.2, and $S$ is less than 2KB in mice-only scenario \cite{vamanan2012deadline}. Moreover, $K$ is set to 24.47KB and $K+K_0$ equals 48KB for each ingress port according to the configuration in \cite{zhu2015congestion}. Generally, as commercial switch owns 32 ingress ports, there are $n=32$. Substituting these values into equations (\ref{PFC}) and (\ref{PFC1}), we can get the probability of triggering PFC at one ingress port is $5.61\times10^{-5}$, and simultaneously at $j$ ingress ports is $e^{-12.235\times32^{j-1}+24\times\frac{32^{j-1}-1}{31}}$. Obviously, these two probabilities are both almost equals zero for any $j$. It means that PFC is rarely triggered and the congestion tree will not grows.
This interprets the good result illustrated in \emph{Fig.}\ref{fig:pfc1-latency} and \emph{Fig.}\ref{fig:pfc2-latency}.
In summary, we have the following corollary.

\noindent\textbf{Corollary 1:}
PFC is enough in mice-only scenario, as it is rarely triggered and hardly incurs the large-scale head-of-line blocking.

On the other hand, according to equations (\ref{PFC}) and (\ref{PFC1}), the probability of triggering PFC  increases with the growth of flow size.
Thus, in the elephant-only scenario, where the flow size is large, the head-of-line blocking issue of PFC becomes significant.
Furthermore, since traffic load of elephant flows are always much heavier than that of mice flows, the head-of-line blocking issue of PFC is aggravated.
Actually, the link utilization can be considered as 100\% in long-term under the elephant-only scenario, and accordingly $P\{Q>n^{j-1}K+\sum_{i=1}^{j-1}n^{i-1}(K+K_0)\}\equiv1$, no matter what other parameters are.
Due to the heavy head-of-line blocking, the link utilization degrades as illustrated in \emph{Fig.}\ref{fig:pfc2-throughput}.
In summary, we have the following corollary.

\noindent\textbf{Corollary 2:}
PFC suffers from heavy head-of-line blocking problem in the elephant-only scenario.

Besides, when only the PFC mechanism is enabled, the elephant flows have negative impact on mice flows. The reason is straightforward: elephant flows leads to frequently PFC triggering, incur head-of-line blocking, and thus block a large number of innocent mice flows.
This interpret the results of \emph{Fig.}\ref{fig:pfc1-latency}.

\subsection{End-to-end congestion control (QCN)}

Here we disable PFC and enable QCN, so as to study the impacts of QCN in the performance impairment shown in \S3.
Similarly, both simulation and theoretical analyses are conducted.

\subsubsection{Simulation}

Repeat the simulation in \S3 with only QCN, the simulation results are shown in \emph{Fig.}\ref{fig:qcn1}.
Compared with the mixed condition, QCN can ensure higher throughput in the elephant-only condition.
This is because the queue length severely oscillates in the mixed condition, as illustrated in \emph{Fig.}\ref{fig:qcn1-queue}. Corresponding to the queue oscillation, there would be packet losses. 
In fact, the ratio of packet loss is about 4.9\% in mixed condition, and quite a number of dropped packets belong to mice flows.
It means the burst mice flows do harm to QCN. 
On the other hand, compared with the mixed condition, QCN can ensure small latency in the mice-only condition, although the queue oscillates severely in both conditions.
This may because the sending rate of mice flow is also limited in the rate adjustment of QCN.
To further verify these understanding, we do theoretical analyses on QCN subsequently. 

\subsubsection{Analyses}
First, observing the small size of mice flows, we realize that QCN has few control on mice flows.
This because the feedback loop delay of QCN is always larger than the transmission delay of mice flows.
For example, assume the size of a mice flow is 2KB. It only needs $1.64\mu s$ to complete transmission of this flow at a $10Gbps$ link. This value is much less than QCN feedback loop delay, which is about $20\mu s$\cite{benson2010network}. 
Therefore, before the feedback of QCN arrives at the source, the mice flow has already been transmitted.
In conclusion, 

\noindent\textbf{Corollary 3:}
QCN is unable to control the mice flow.

\begin{figure}[t]
	\centering
	\subfigure[Latency of mice flows]{
		\includegraphics[width=0.45\linewidth]{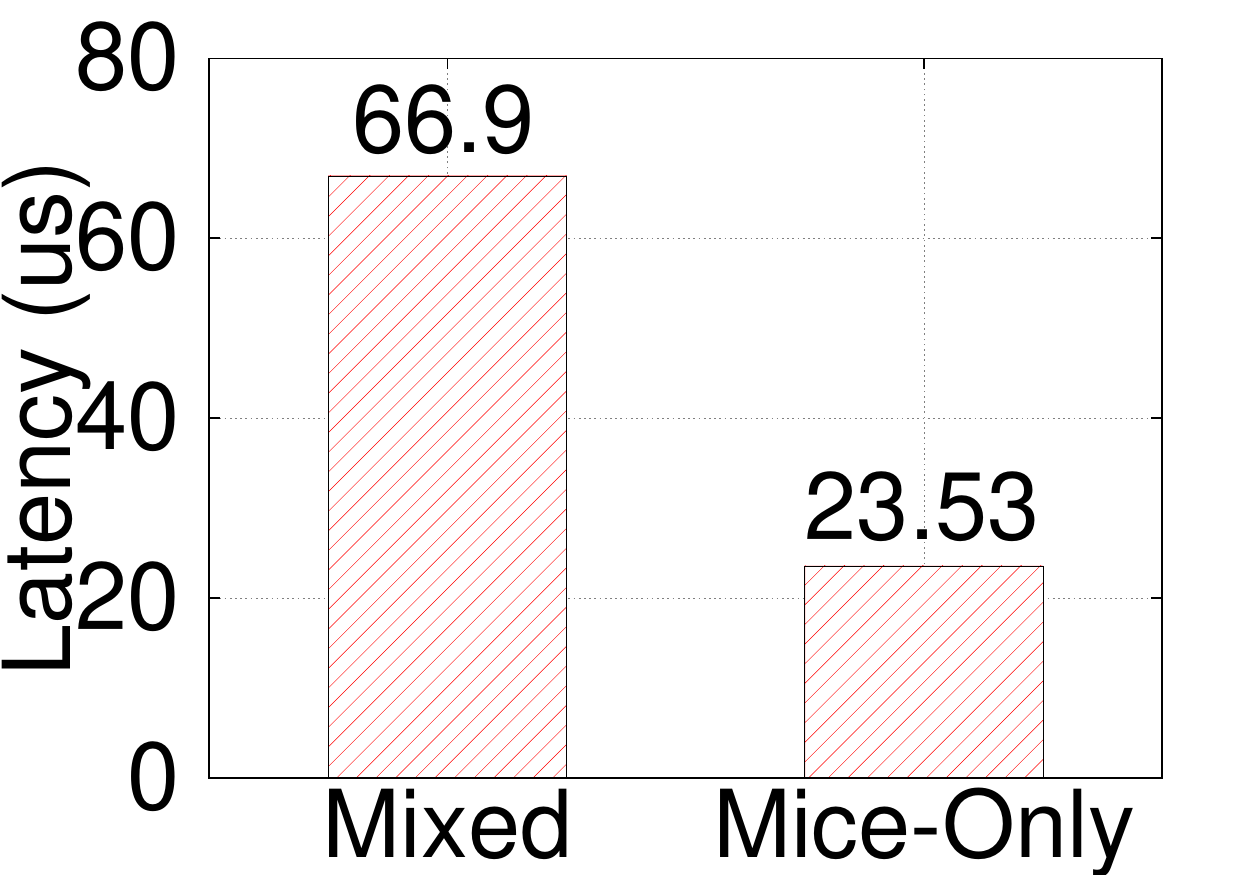}
		\label{fig:qcn1-latency}
	}
	\hfill
	\subfigure[Bottleneck throughput]{
		\includegraphics[width=0.45\linewidth]{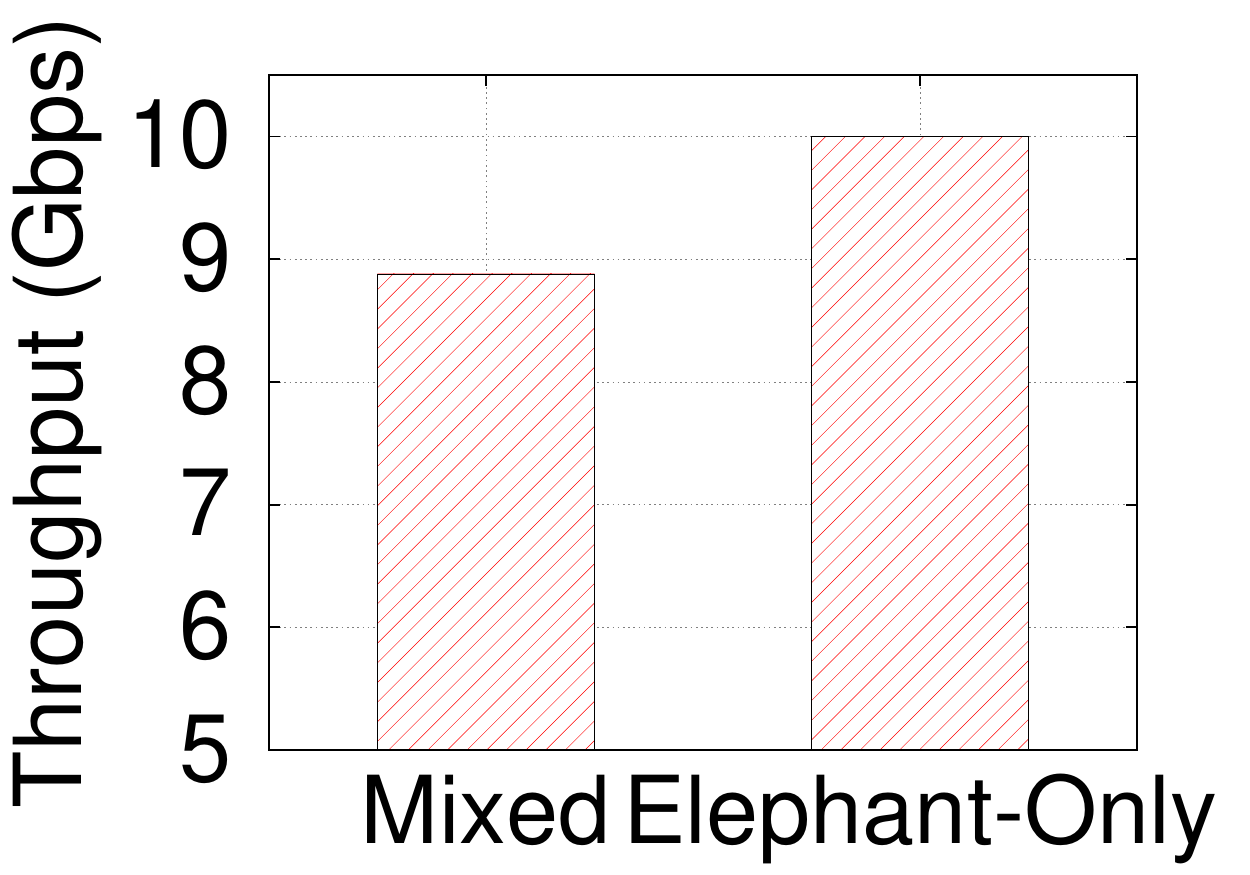}
		\label{fig:qcn1-throughput}
	}
	\subfigure[Bottleneck queue length]{
		\includegraphics[width=1\linewidth]{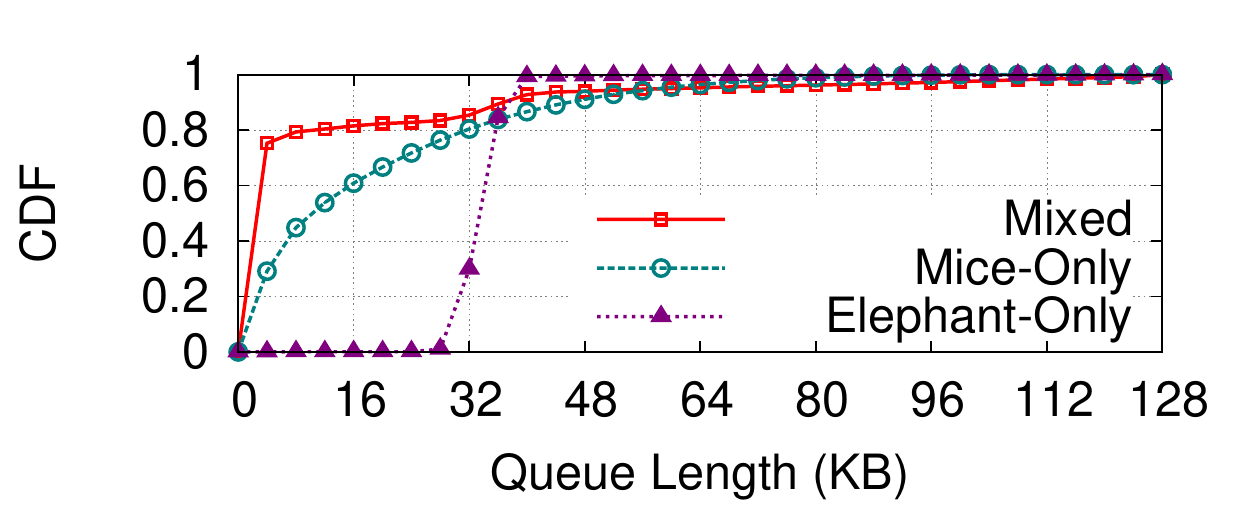}
		\label{fig:qcn1-queue}
	}
	\subfigure[Occupations of mice when QCN is triggered.]{
		\includegraphics[width=0.45\linewidth]{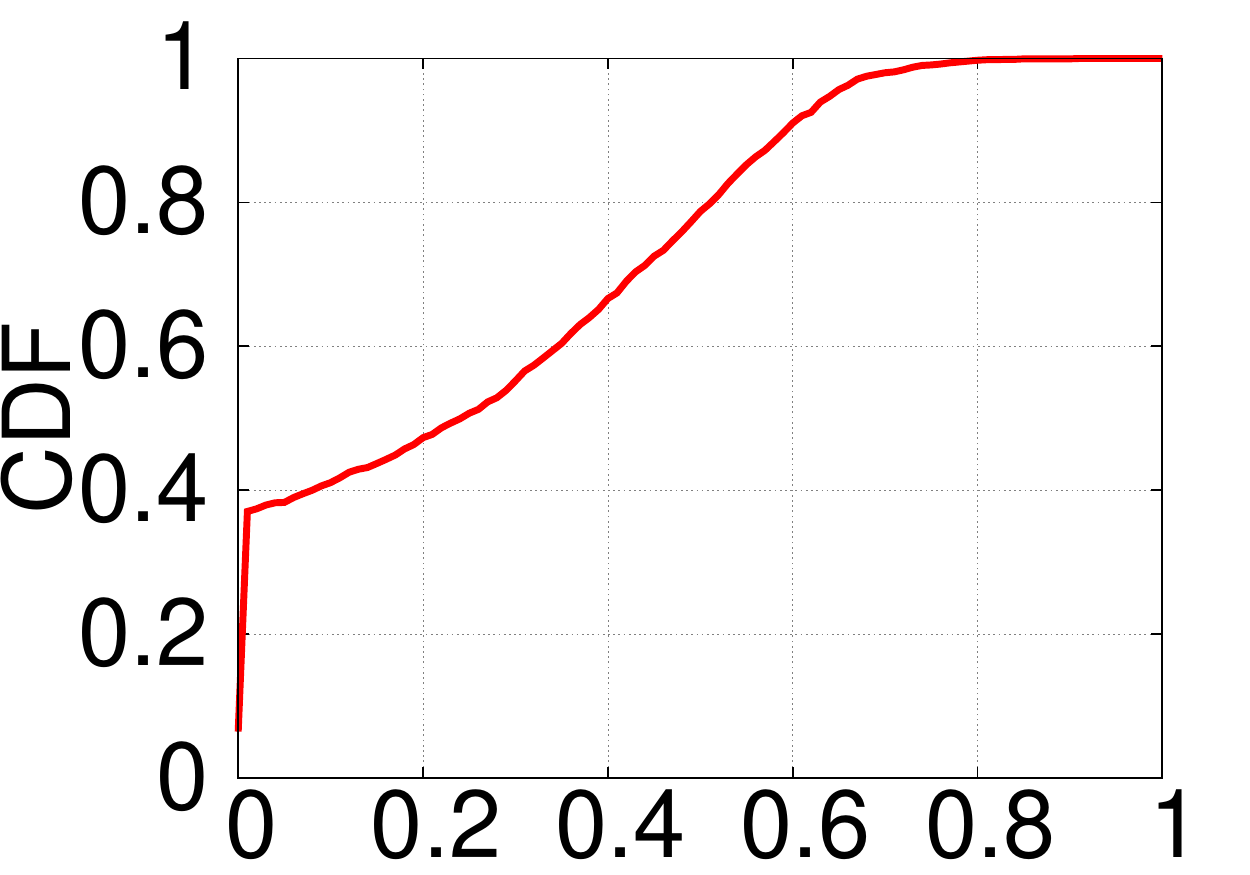}
		\label{fig:qcn1-short}
	}
	\subfigure[Feedback value]{
		\includegraphics[width=0.45\linewidth]{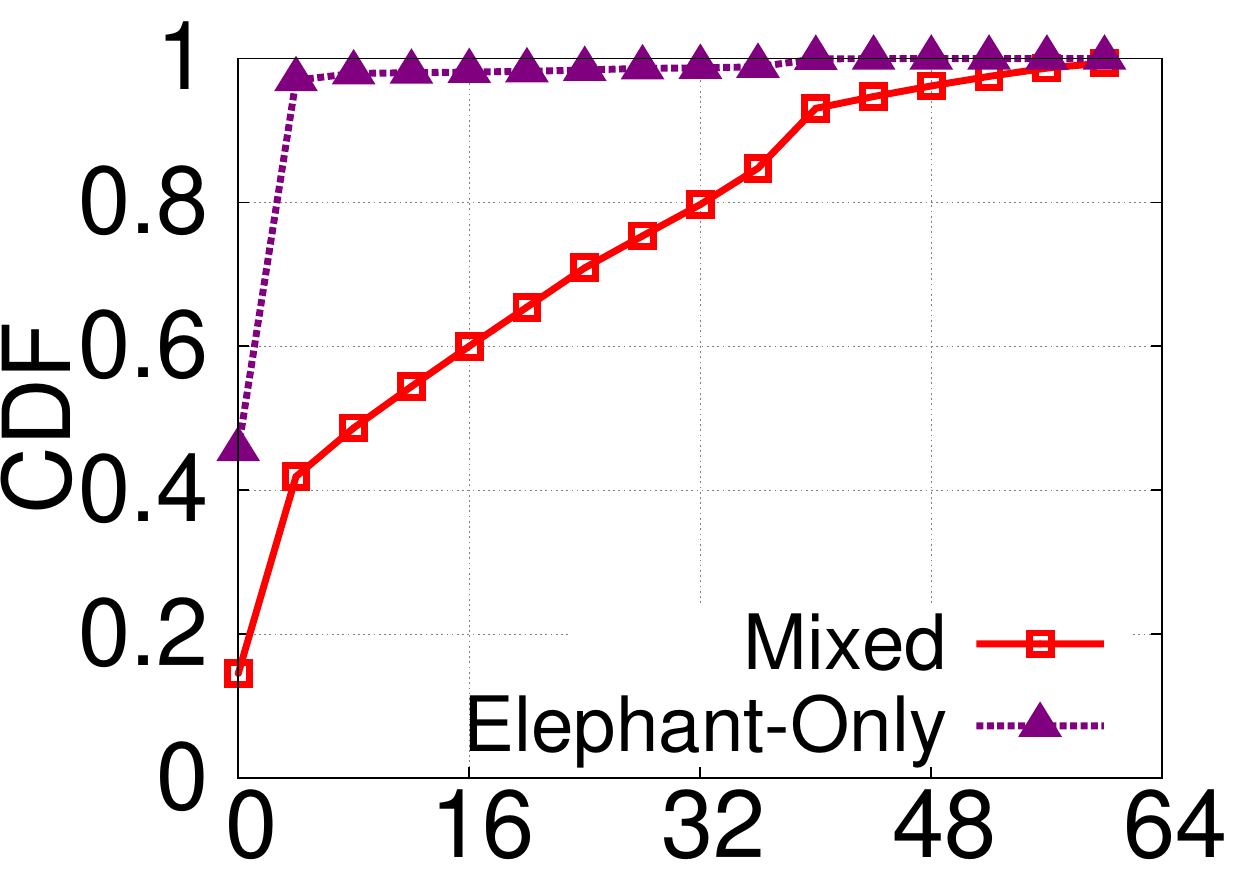}
		\label{fig:qcn1-fb}
	}
	\caption{QCN in the many-to-one scenario.}
	\label{fig:qcn1}
\end{figure}

With this insight, we can interpret simulation result \emph{Fig.}\ref{fig:qcn1-latency}. In the mice-only condition, the mice flows are transmitted freely. 
While in the mixed condition, the sending rate of mice flows is victimized when QCN limits the sending rate of elephant.

Furthermore, the burst mice flows generate invalid feedback to QCN. This point can be classified by calculating the probability of rate decrease in QCN. 
Assume the congestion control system is in stable state, where $Q_{off}=0$ and $Q_\delta=0$. Once $Q_\delta >0$, CP will tell the source to slow down. Let $\Delta R$ represent the transient queue length growth rate, $\tau$ be the stochastic interval between two successive flows and $F_\tau(x)$ denote the distribution function of $\tau$.
Obviously, $Q_\delta>0$ is equivalent to $\Delta R>0$ on the respect of triggering rate decrease. 
Similar to \S4.1, we can define the random flow sizes $S$ and average flows arrival rate $\lambda$. 
There are
\begin{equation}
\begin{array}{cl}
	P\{\Delta R>r\}&=P\{\tau <\frac{S}{r}\} \\[5pt]
	& =F_\tau\left(\frac{E(\tau)\lambda}{r}\right)=F\left(\frac{\lambda}{r}\right)
\end{array}
\label{eq:qcn}
\end{equation}
Equation (\ref{eq:qcn}) indicates that the more burstiness (heavy tailed $F(x)$) or larger workloads (larger $\lambda$) lead to the more intensive rate decrease and the lower link utilization. 
The burstiness is right the part that mice flows play, but the corresponding rate decreases act on elephant flows.
While the mice flows are not effected according to corollary 3.
In fact, the average occupation of mice when QCN is triggered is about 25\%, and the bursty short flows increase the feedback value for long-lived flows about eight times, as shown in \emph{Fig.}\ref{fig:qcn1-short} and \emph{Fig.}\ref{fig:qcn1-fb}.
In total, we have

\noindent\textbf{Corollary 4:}
QCN is interfered by the burst mice flow, which leads to the generation of many invalid feedback.

\subsection{Discussion}
\begin{figure}[t]
	\centering
	\subfigure[Latency of mice]{
		\includegraphics[width=0.45\linewidth]{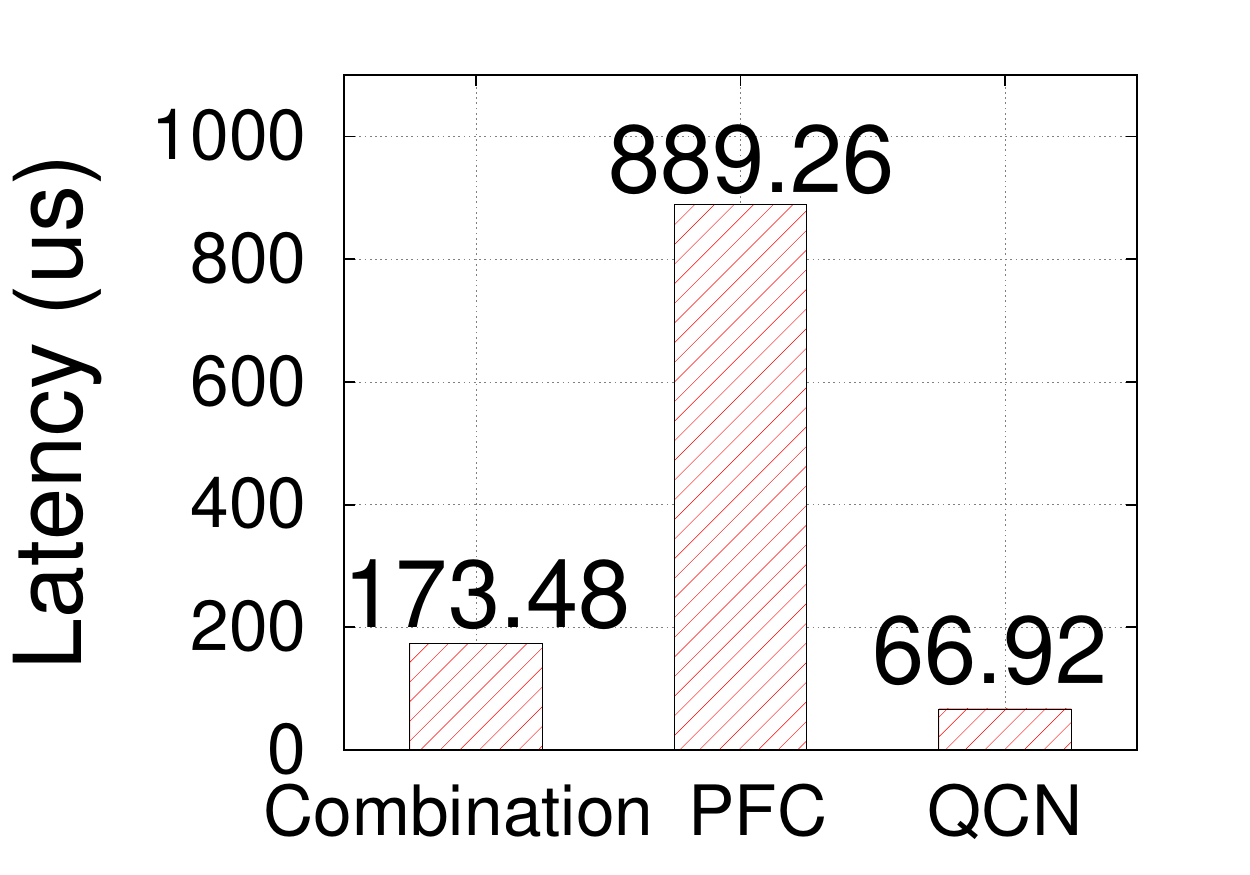}
	}
	\hfill
	\subfigure[Bottleneck throughput]{
		\includegraphics[width=0.45\linewidth]{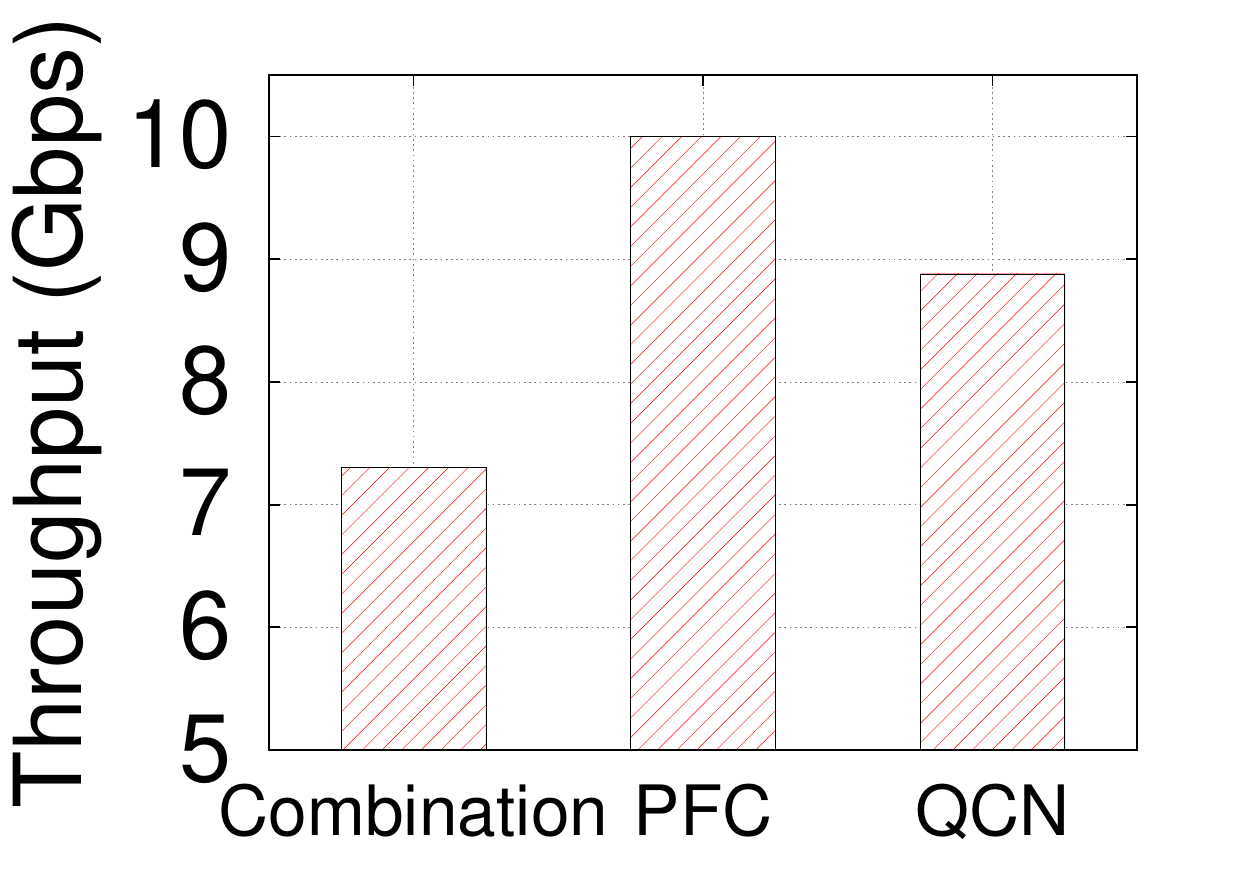}
	}	
	\caption{Combination of PFC and QCN performances worse than PFC-only and  QCN-only scenarios.}
	\label{fig:worse}
\end{figure}

\begin{figure}[t]
	\centering
	\includegraphics[width=0.95\linewidth]{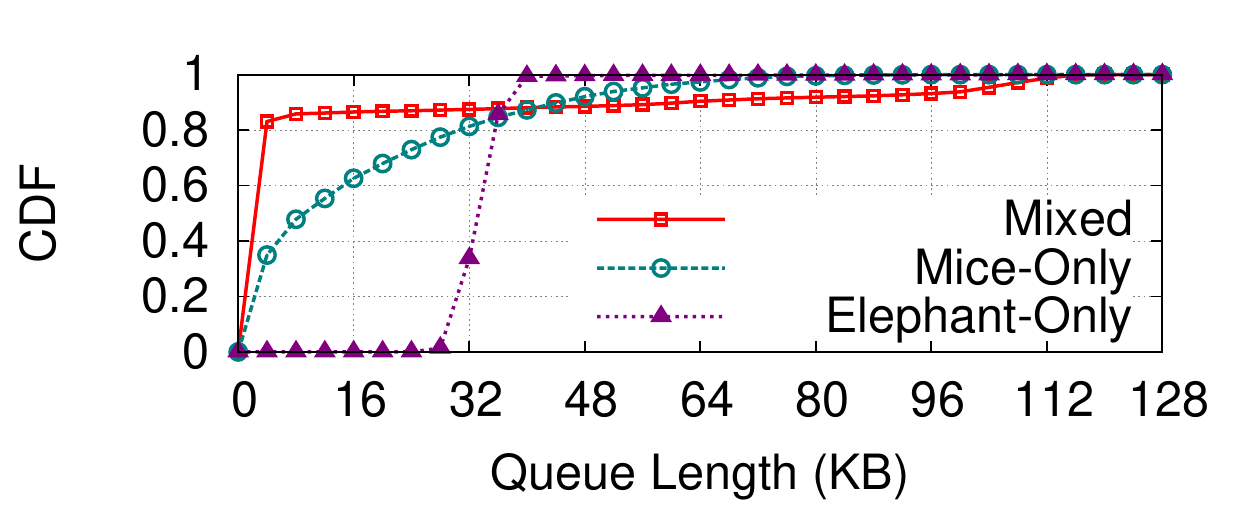}
	\caption{Bottleneck quque length}	
	\label{fig:m_queue}
\end{figure}

\begin{figure}[t]
	\centering
	\subfigure[Number of PAUSE events]{
		\includegraphics[width=1\linewidth]{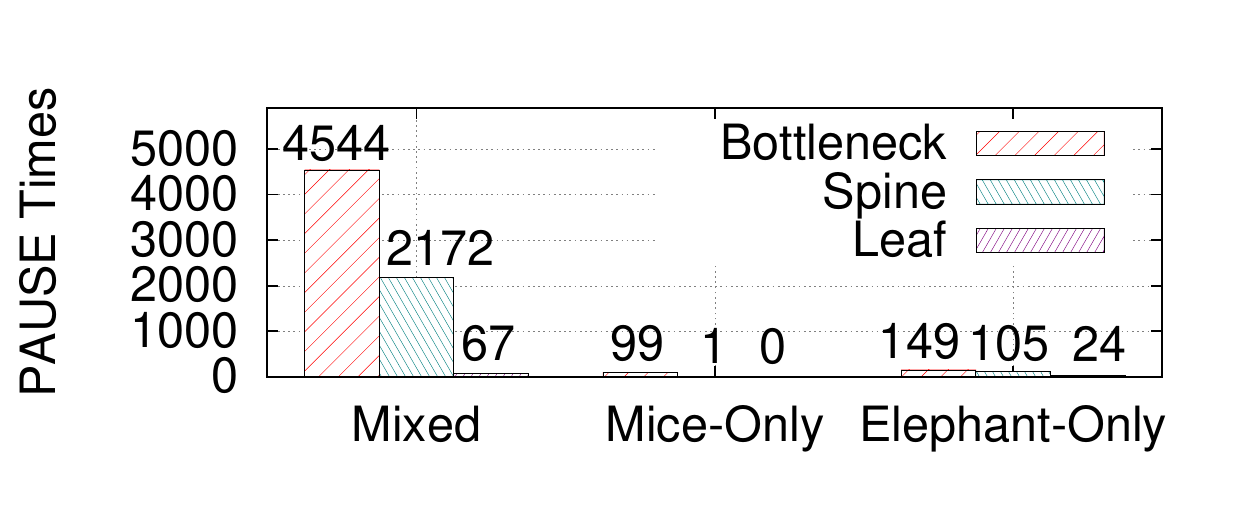}
		\label{fig:m_pause}		
	}
	\subfigure[Feedback value]{
		\includegraphics[width=1\linewidth]{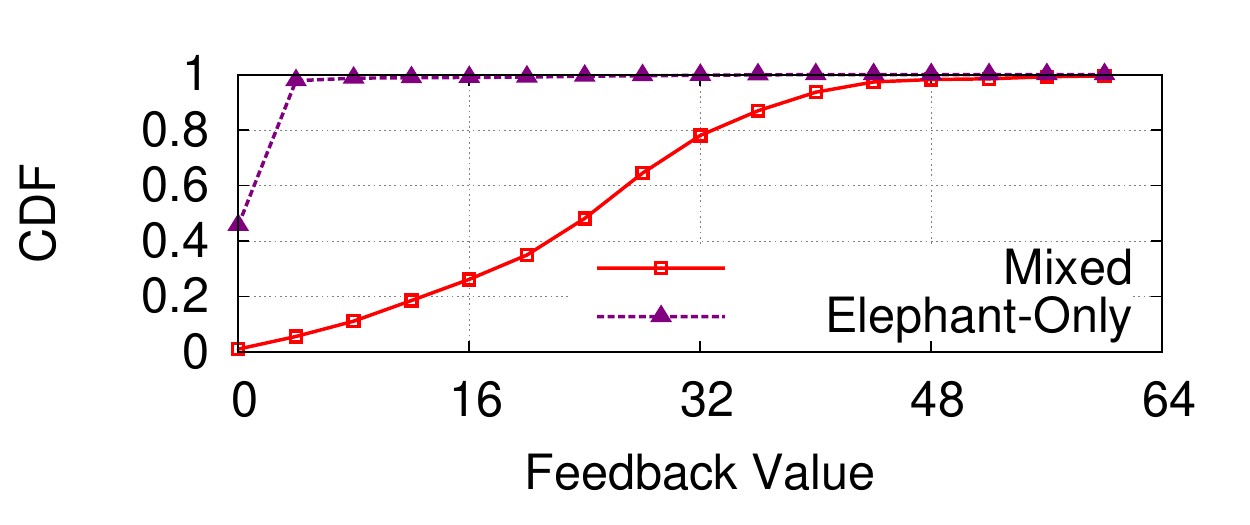}
		\label{fig:m_fb}		
	}
	\subfigure[Occupations of elephant when PFC is triggered.]{
		\includegraphics[width=0.45\linewidth]{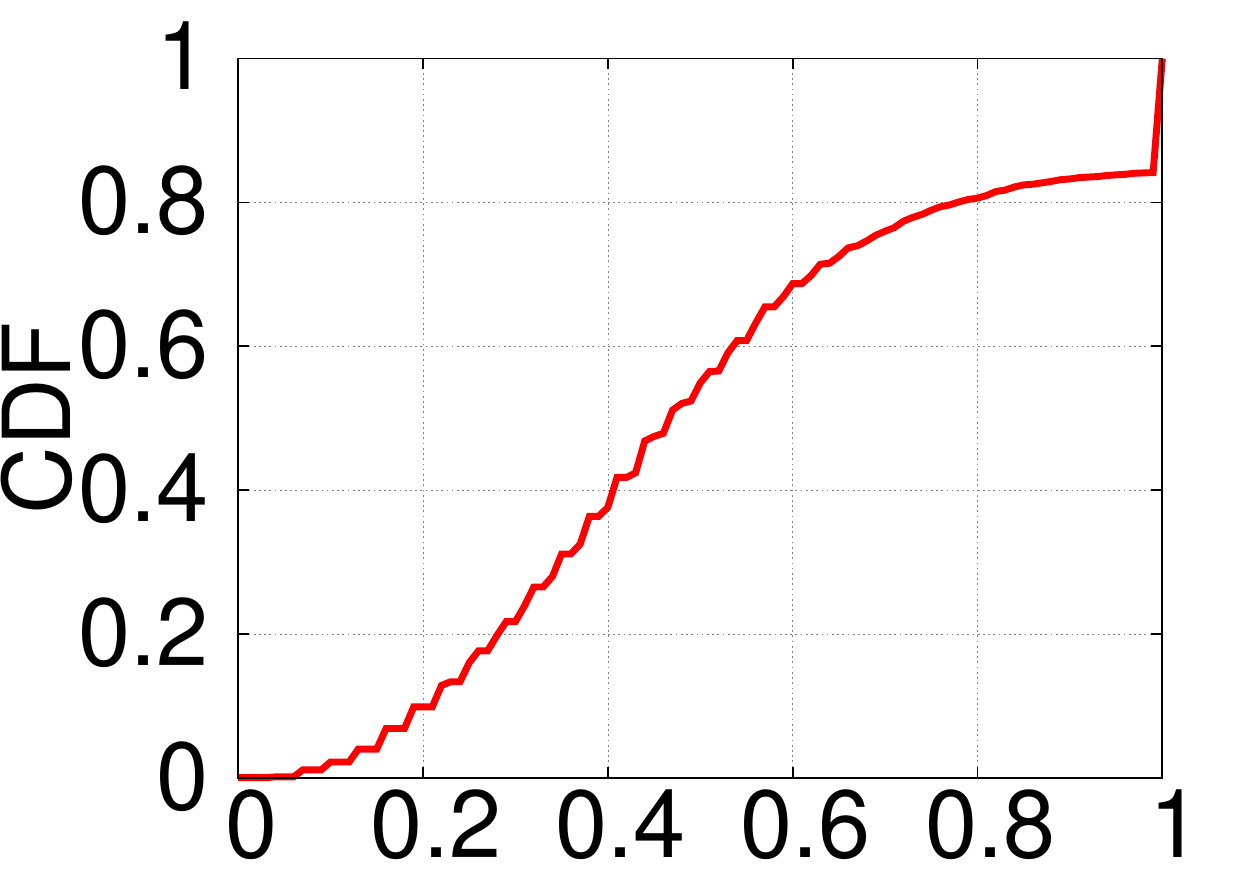}
		\label{fig:long}
	}
	\hfill
	\subfigure[Occupations of mice when QCN is triggered.]{
		\includegraphics[width=0.45\linewidth]{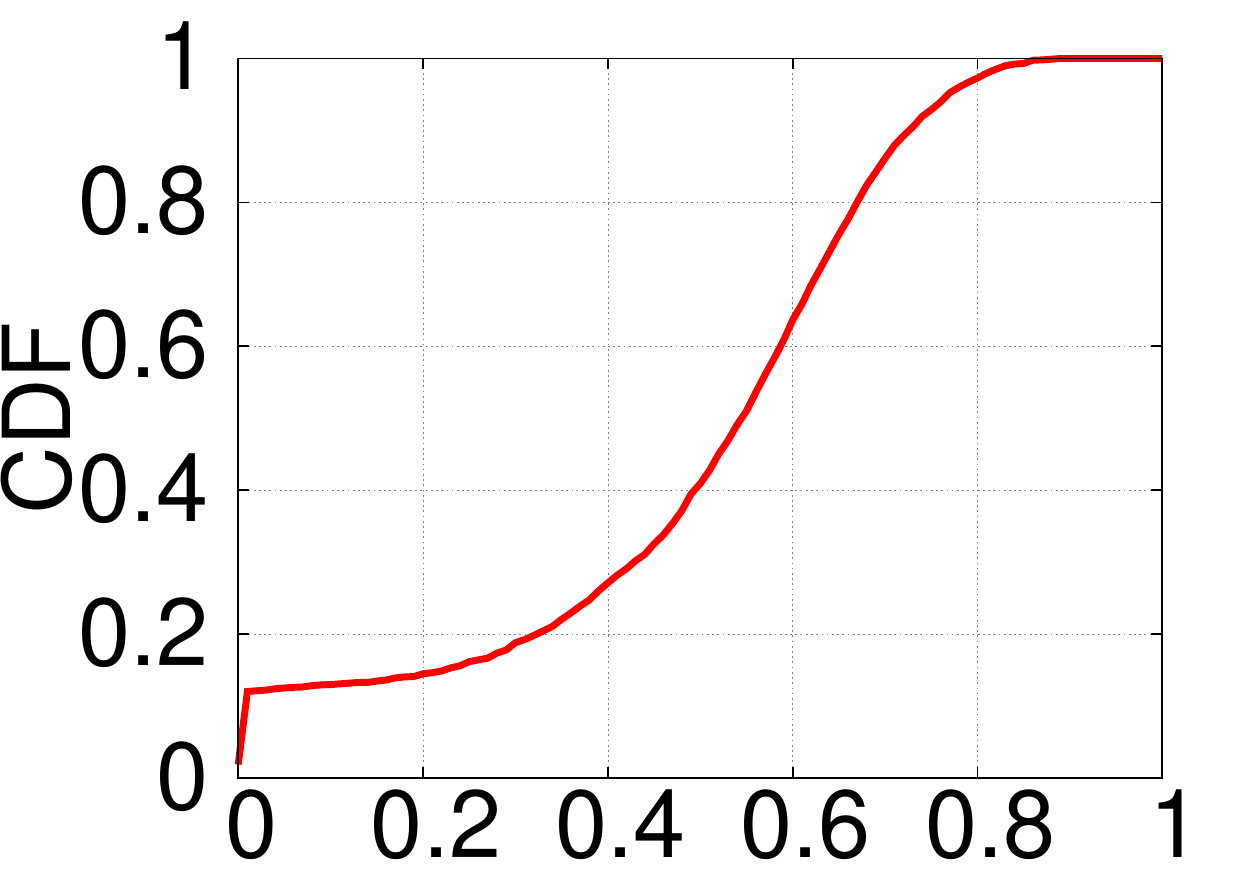}
		\label{fig:short}
	}
	\caption{Interaction of mice and elephant flows with combination of PFC and QCN.}
\end{figure}
The analysis above implies that PFC is good enough for mice and QCN is appropriate for elephant, i.e. the hop-by-hop and end-to-end flow control mechanisms can handel transient and persistent congestions respectively.
However, the combination of these two flow control mechanisms do not helps well with mixture traffic, as shown in \S\ref{motivation}.
Overview on our nine simulations of  many-to-one, we are surprised that the performance of mixed traffic under the combination of PFC and QCN is not only worse than that in the mice-only and elephant-only scenarios, but also worse than that in the PFC-only and  QCN-only scenarios.
As shown in \emph{Fig.}\ref{fig:worse}, although the mice latency in combination scenario is quite smaller than that in PFC-only scenario, it is still as large as triple of the latency in QCN-only scenario.
And the impairment of bottleneck throughput is obvious comparing with the single flow control scenario.

Furthermore with the combination of PFC and QCN shown in \emph{Fig.}\ref{fig:m_queue}, the bottleneck queue length oscillates frequently but not severely in mice-only scenario, and converges directly without oscillation in elephant-only scenario, but oscillates severely in mixed scenario,  causing a large number of PAUSE events and link underutilization.
We also count the total number of PAUSE messages generated from bottleneck switch Leaf 4 and spreading in the network. As shown in \emph{Fig.}\ref{fig:m_pause}, the mixed traffic triggers much more PAUSE events than the mice-only or elephant-only scenarios.
And the feedback for elephant im mixture traffics mainly values in $[16\sim32]$, while rarely grows above $2$ in elephant-only scenario.
Finally, the occupation of elephant when PFC is triggered and  occupation of mice when QCN is triggered are shown in \emph{Fig.}\ref{fig:long} and \ref{fig:short}.
In average these occupations are both about 50\% while more than 15\% PAUSE events and feedback generation are caused when the bottleneck queue is full of elephant.

\begin{table}[!tp]
	\centering
	\begin{tabular}{|c|c|c|c|}
		\hline
		& Mice & Elephant  & Mixture  \\
		\hline
		PFC & Good & Bad & Bad \\
		& & Congestion tree & Congestion tree\\
		\hline
		 & Bad  &  & Bad\\
		QCN & Loss & Good & Loss,Queueing \\
		& & & Throughput loss \\
		\hline
		PFC &   &  & Bad \\
		and & Good & Good & Queueing \\
		QCN & & & Throughput loss \\
		\hline
	\end{tabular}
	\caption{Hop-by-hop and end-to-end flow control mechanisms with the mixture of mice and elephant.}
	\label{tab:dcb}
\end{table}
The key findings of our simulations are summarized in Table \ref{tab:dcb}.
We conclude the root cause of the performance impairment as that the congestion detection signals are polluted when the hop-by-hop flow control and the end-to-end flow control interacts in one queue.
The overshoot in the regulation process of end-to-end flow control system likely touch the queue threshold of hop-by-hop flow control scheme. On the other side, frequent short flow arriving is noise to disturb the end-to-end flow control correctly judge persistent congestion status.

\section{Design}
\label{solution}
\subsection{Principle}
\label{principle}
\begin{figure}[t]
	\centering
	\includegraphics[width=0.9\linewidth]{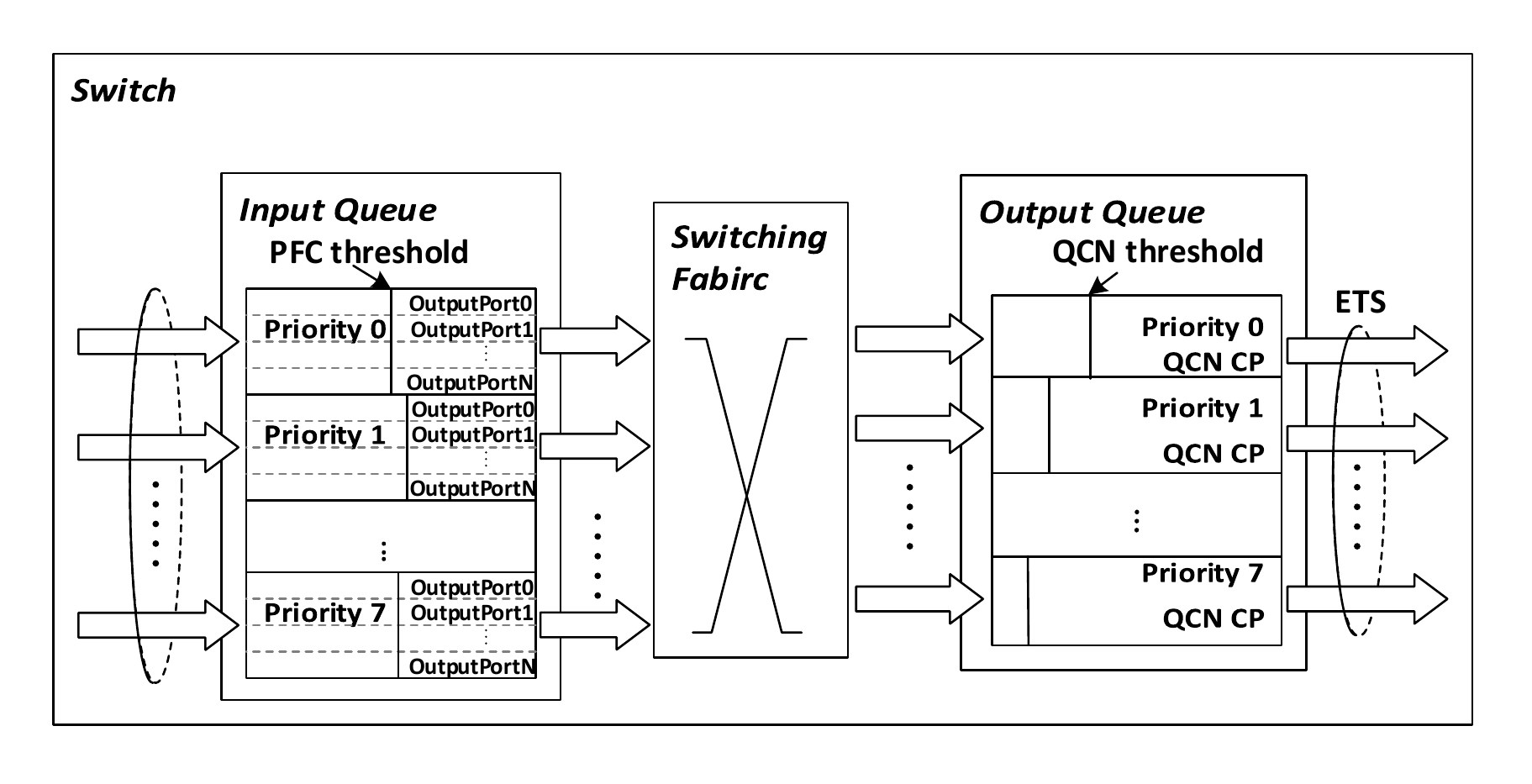}
	\caption{DCB  already provides a proposal that isolates different traffics.}
	\label{fig:dcb}
\end{figure}
As discussed in \S\ref{understand}, either the hop-by-hop flow control scheme PFC or the end-to-end congestion control scheme QCN fails to simultaneously satisfy the requirements of both mice and elephant, and so does the combination of them.
While PFC fits the mice-only scenario and QCN adapts to the elephant-only scenario.
The root cause of these failures are concluded due to the the congestion detection signals are polluted when the hop-by-hop flow control and the end-to-end flow control interacts in one same queue.
Naturally, we isolate the short-lived and long-lived flows and assign the appropriate flow control scheme for each, we can perhaps achieve the ideal performance in the mixture scenario.
This idea is analogous to the road traffic control system in a city where buses and cars are separated into different lanes and follow the different traffic law.

The contributions of isolation are as follows,

(1) The long flows would not introduce unnecessary PAUSE events harming the short flows.
	For example, if we assign a high priority for mice flows and a low one for elephant, mice flows will need not wait for the queue built by elephant flows to drain and will rarely be paused. 
The reduction of pause events avoid the harms of congestion-spreading and unfairness issue of PFC thus we can get better completion time for short flows and higher link utilization.

(2) The bursty arrivals of short flows are mitigated for long flows. That is, the long flows will only feel that the short flows arrive at the rate not larger than  the bandwidth.
Consequently, long flows experience a lighter rate-decrease and the switch occupation   oscillates more smoothly avoiding link utilization-loss due to empty buffer and frequent packet-drop.


Evaluation results in \S\ref{evaluation} verify the great improvement of our isolation proposal that the completion time of short flows is reduced by 85\% and the throughput of elephant are increased by 20\%.

\subsection{Deployment}
The DCB architecture has already provided mechanisms that can be used to isolate different traffics, e.g. Priority Groups or Enhanced Transmission Selection.
DCB isolates traffics in different priority groups according to their own traffic types such as LAN, SAN, HPC, etc (shown in Table~\ref{tab:class}).
Each traffic type is mapped into an exclusive virtual interface and obtains the bandwidth allocated for it.
As shown in \emph{Fig.}\ref{fig:dcb}, flow control schemes like PFC and QCN can be locally set and independently operates among different priority groups.
\begin{table}[h]
	\centering
	\begin{tabular}{|c|c|c|}
		\hline
		Application & User Priorities & Bandwidth Share  \\
		\hline
		LAN  Traffic & 0,3,4,5 & 40\%  \\
		\hline
		SAN  Traffic & 1,6 & 40\%  \\
		\hline
		HPC  Traffic & 2,7 & 20\% \\
		\hline
	\end{tabular}
	\caption{DCB isolates traffics in different priority groups according to their traffic types.}
	\label{tab:class}
\end{table}

Note that there exist both mice and elephant flows in the same traffic type and each type has more than one user priorities. In this context, we can isolate the mixed flows by these priorities within a single traffic type without affecting other types. On the basis of the isolated traffics, we can assign the corresponding suitable flow control schemes for each traffic pattern, e.g. PFC for mice flows and QCN for elephant flows.

\subsection{Parameter Settings}
Follow the DCB architecture, we can deploy our isolating proposal in current devices readily.
The main configuration parameters are two-fold.
\subsubsection{Buffer allocation }
For mice flows, we need to guarantee a large enough PFC threshold, or PFC will pause many hops and cause head-of-line blocking as Eq. \ref{PFC} and \ref{PFC1} indicate. However, when it comes to elephant flows, just a small queue length in buffer can guarantee throughput. Therefore, the limited buffer should be preferentially allocated to mice flows and we can simply assign the remaining buffer to elephant flows.

As discussed in \S\ref{understand}, when the PFC threshold $K$ is set to 24.47KB as decided in \cite{zhu2015congestion}, the probability of pausing more than one hop almost equals zero. Moreover, $K_0$--the data transmitting during the pause frame travels to the last hop is only a little smaller than $K$. Since for PFC to work normally, the buffer size in each ingress port should be at least $K+K_0$, here we apportion 48KB of buffer to mice flows.Correspondingly, the remaining buffer is allocated to elephant flows.

\subsubsection{Bandwidth allocation} 

The enhancement of ETS provides a simple bandwidth allocation method for all the priority groups.
In \S\ref{principle}, we choose a scheduler algorithm of strict priority   that mice take the higher priority and elephant has to wait to be scheduled when mice are arriving.
Thus the noise to the long-lived flows caused by frequent short flow arriving is fixed in the level of $C$ rather than the real high speed.
Thus once we remains a part of bandwidth for elephant, for example 10\%, the impacts from mice on elephant can be controlled at the level of $0.9C$, which would improve elephant throughput.
However, on the other hand, remaining some bandwidth for elephant implies that mice need to wait in a violent bursty arrival. 
We will verify the impact of bandwidth allocation in \S\ref{evaluation}.


\begin{figure}[t]
	\centering
	\subfigure[Latency of mice] {
		\includegraphics[width=0.45\linewidth]{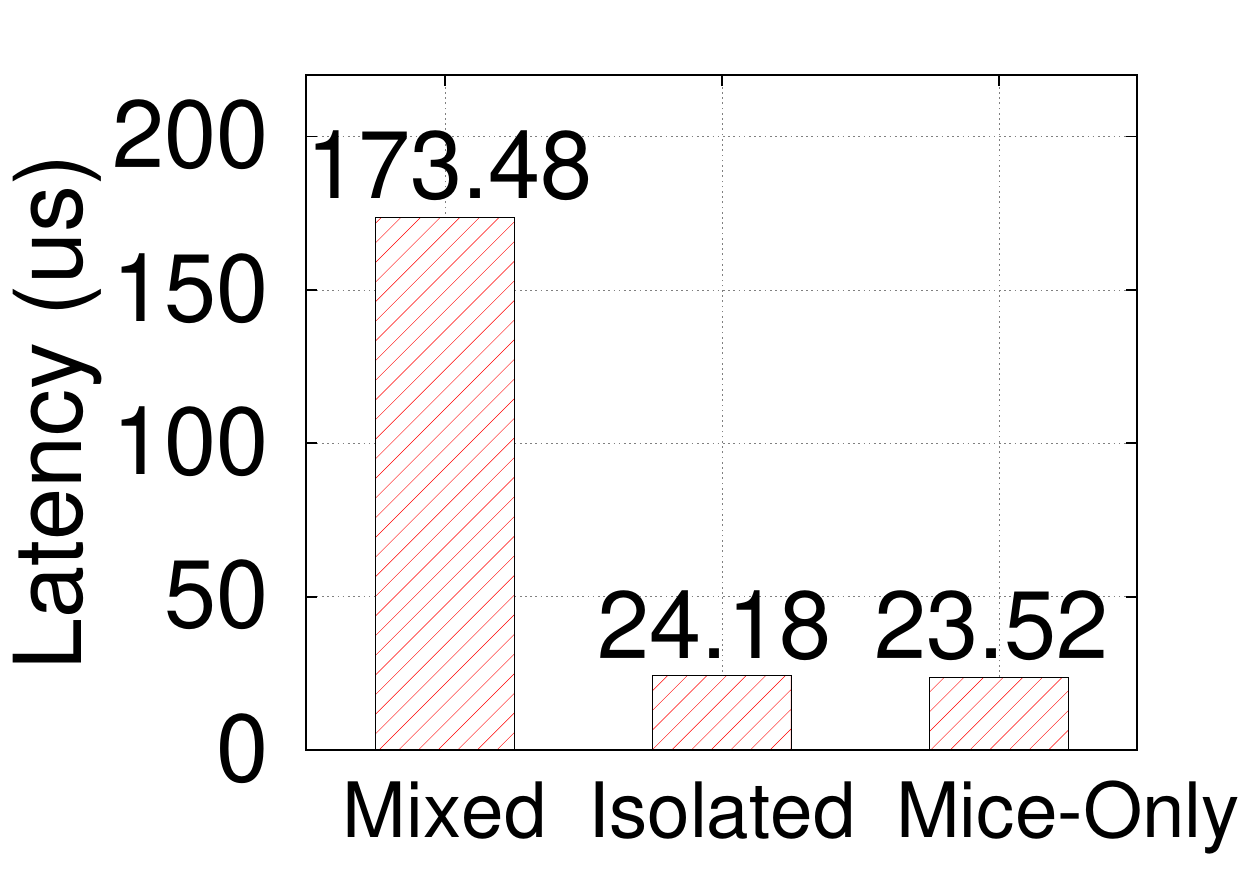}
		\label{fig:s_latency}
	}
	\hfill
	\subfigure[Bottleneck throughput] {
		\includegraphics[width=0.45\linewidth]{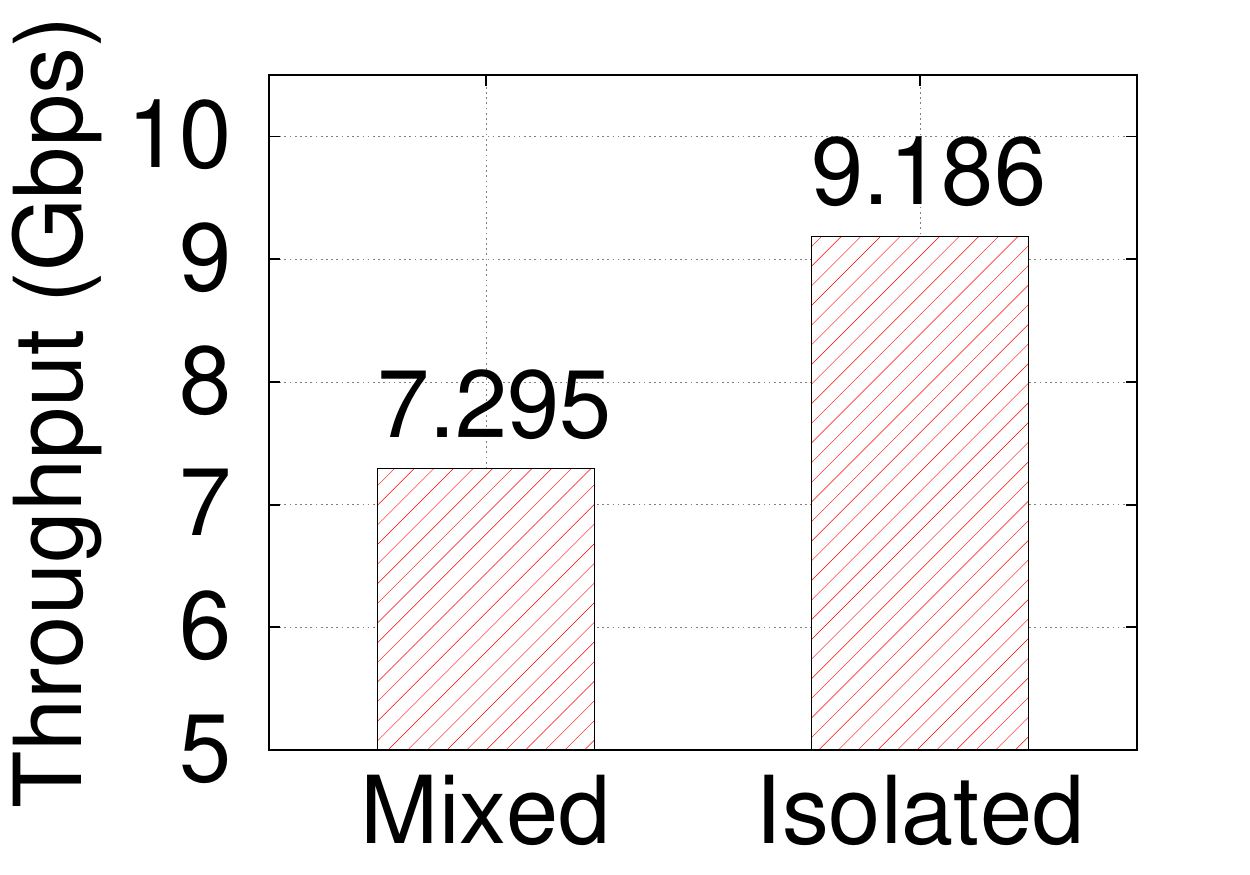}
		\label{fig:s_throughput}
	}
	\caption{Isolation improves latency for mice and throughput for elephant. }
	\label{fig:s}
\end{figure}

\subsection{Discussion}
\subsubsection{Boundary of mice and elephant}
According to \emph{Fig.}4 of \cite{alizadeh2011data}, the numerous short-lived flows are mainly smaller than $10KB$ and the heavy long-lived are almost larger than $100KB$. Any size between $10KB$ and $100KB$ can be a recognized boundary of mice and elephant.
After the boundary is defined, it is fairly convenient for the source to decide the type of current flow, since generally the total length of one flow is available before sending the first packet. Even this information is unavailable, source can set flow as mice at first and changes it into elephant when transmitted size exceeds the boundary.
In fact, the original distinctions between mice and elephant are not only flows size but also requirement.
One application can configure its traffic as mice or elephant based on  empirical data and realistic performance requirements.
\subsubsection{Upper layer end-to-end mechanisms}

In reality, the L2 congestion control mechanism QCN may not be widely appropriate, especially in large-scale data centers where the L2 control frames fail to  travel through the IP-routed networks \cite{zhu2015congestion}.
To address this problem, we extend our isolation proposals to  L4 or L3 congestion control schemes such as TCP and DCTCP, which serves as the end-to-end flow control mechanism for elephant. 
As verified in \S\ref{evaluation}, our proposal of isolation performances similarly no matter selecting any end-to-end flow control scheme for elephant.

\section{Evaluation}
\label{evaluation}
This section is divided in two parts. First,  we validate the benefit of our proposal of isolation.  
Next, we extend our proposal introducing the widely used end-to-end flow control schemes of TCP and DCTCP instead of QCN.
In these two steps, we still use the many-to-one scenario in \S\ref{motivation} and \S\ref{understand} for comparison.
Finally, this isolation proposal is evaluated in a large-scale scenario. 
\subsection{Benefit}

\begin{figure}[t]
	\centering
	\subfigure[Number of Pause events] {
		\includegraphics[width=1\linewidth]{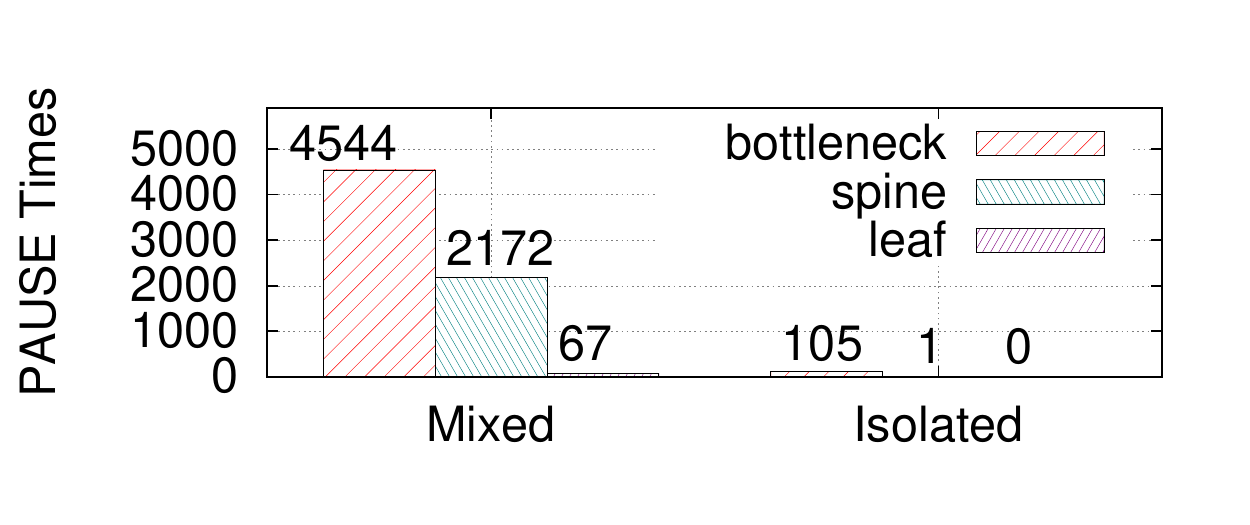}
		\label{fig:s_pause}
	}
	\subfigure[Feedback value] {
		\includegraphics[width=1\linewidth]{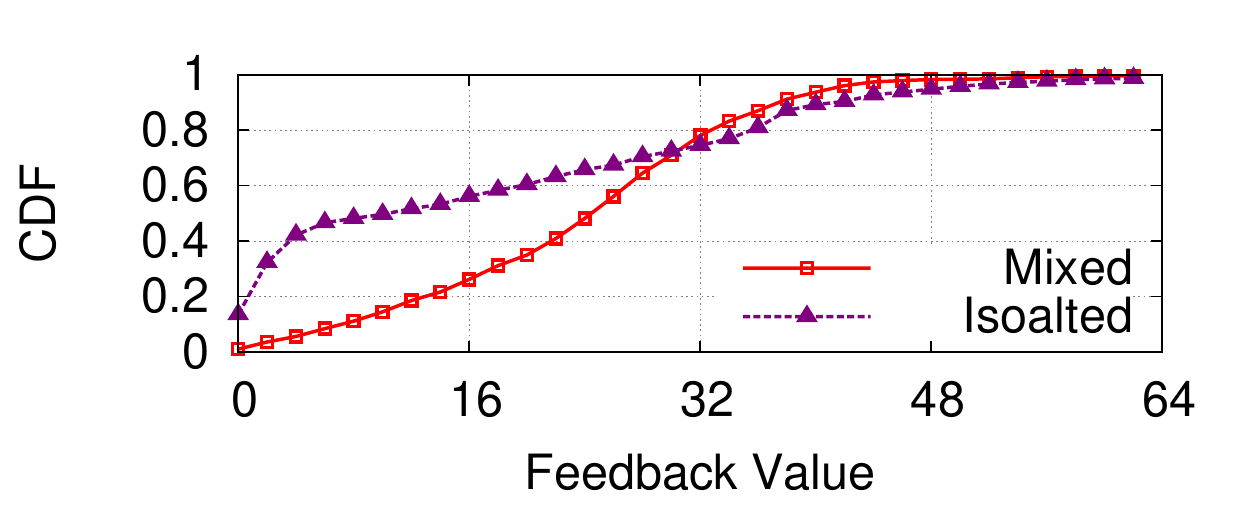}
		\label{fig:s_fb}
	}
	\caption{Isolating mitigates the Interaction of mice and elephant flows with combination of PFC and QCN. }
	\label{fig:solution}	
\end{figure}

Firstly, we assign a high priority for mice and a low one for elephant that the mice would be transmitted while competing with elephant.
As shown in \emph{Fig.}\ref{fig:s}, with this isolation operation, the latency of mice is reduced 85\%, as small as the latency in mice-only scenario and the bottleneck throughput increases about 2Gbps.
Results in \emph{Fig.}\ref{fig:solution} verify our prediction that isolating mice and elephant leads to much fewer pause events for mice and low feedback value for elephant.

\begin{figure}[t]
	\centering
	\subfigure[Latency of mice]{
		\includegraphics[width=0.45\linewidth]{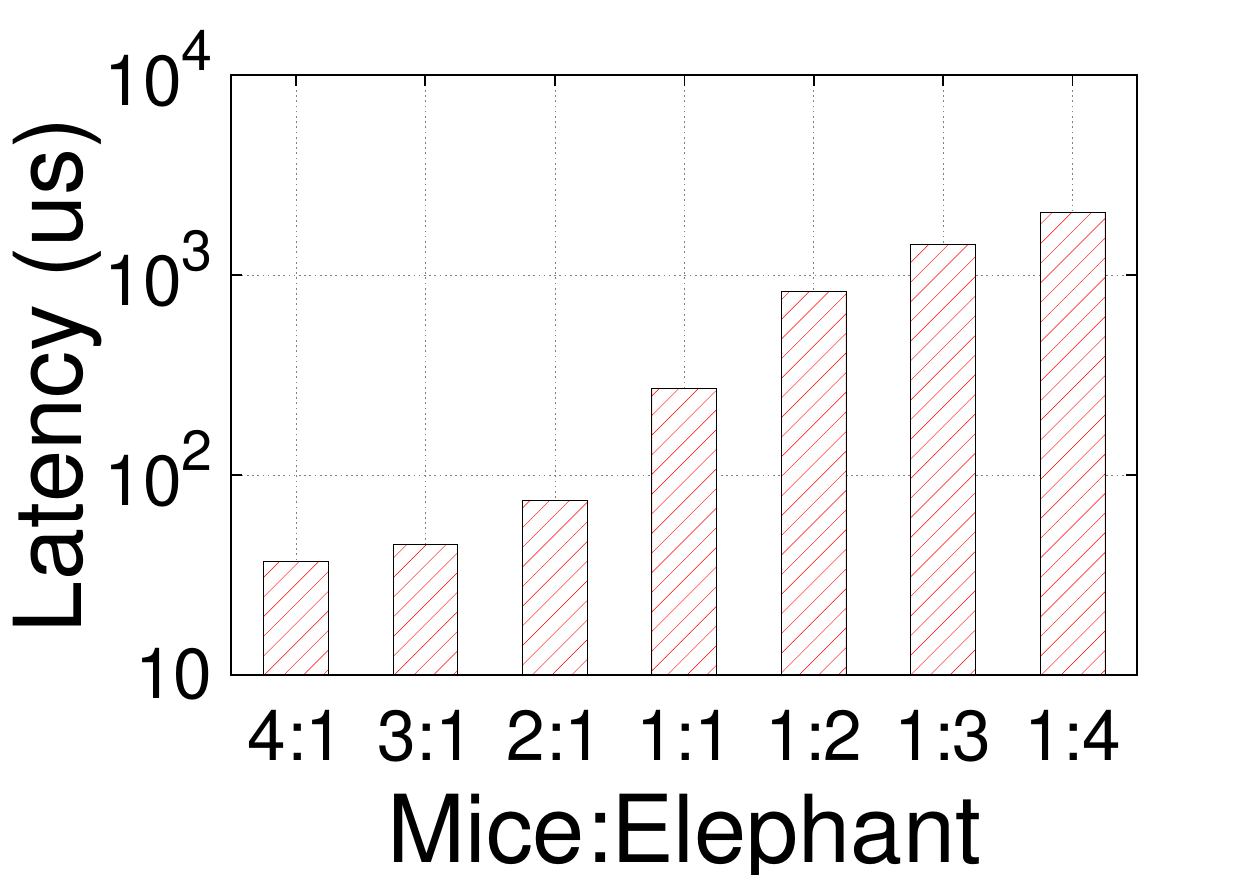}
	}
	\hfill
	\subfigure[Bottleneck throughput]{
		\includegraphics[width=0.45\linewidth]{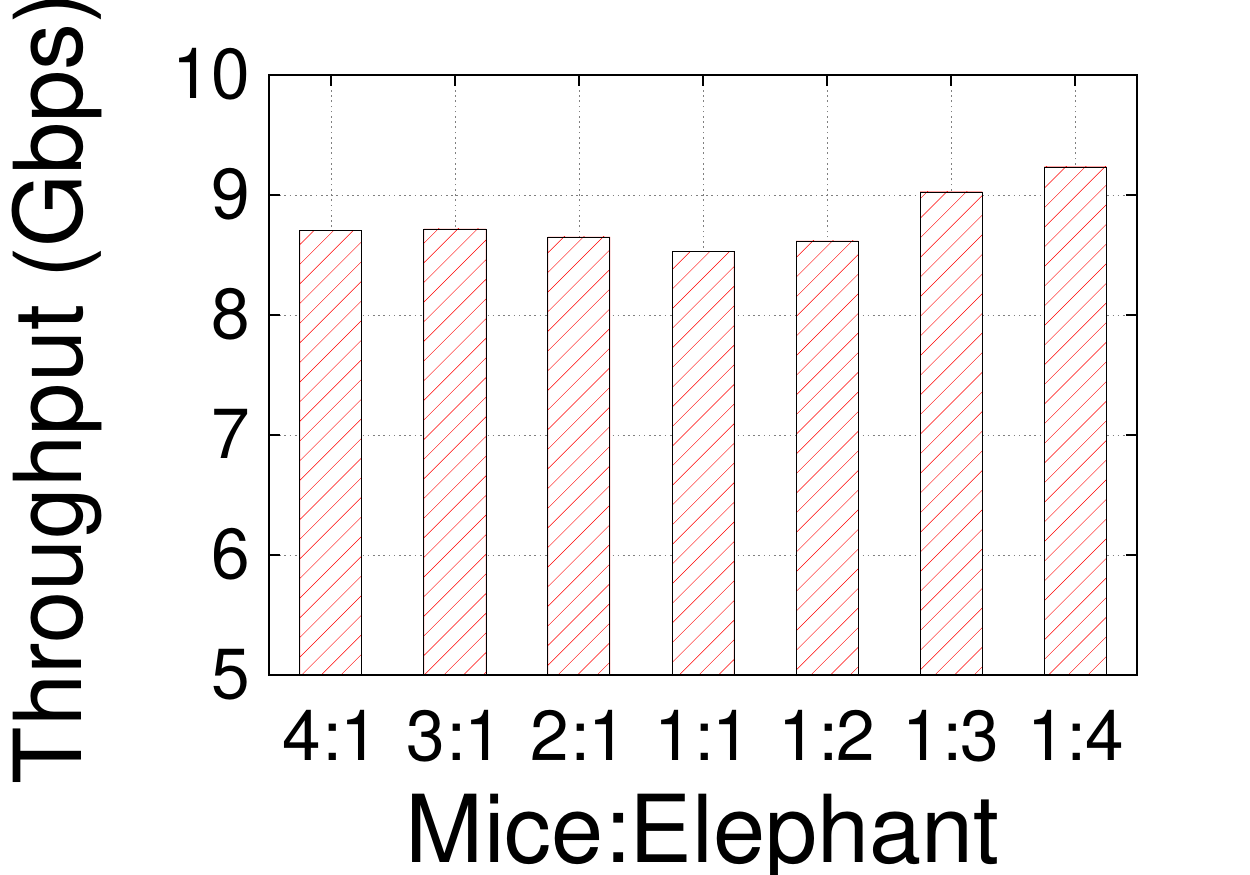}
	}
	\caption{Isolation results while bandwidth allocation of mice and elephant varies. }
	\label{fig:ets}
\end{figure}

\begin{figure}[t]
	\centering
	\subfigure[Number of PAUSE events]{
		\includegraphics[width=0.45\linewidth]{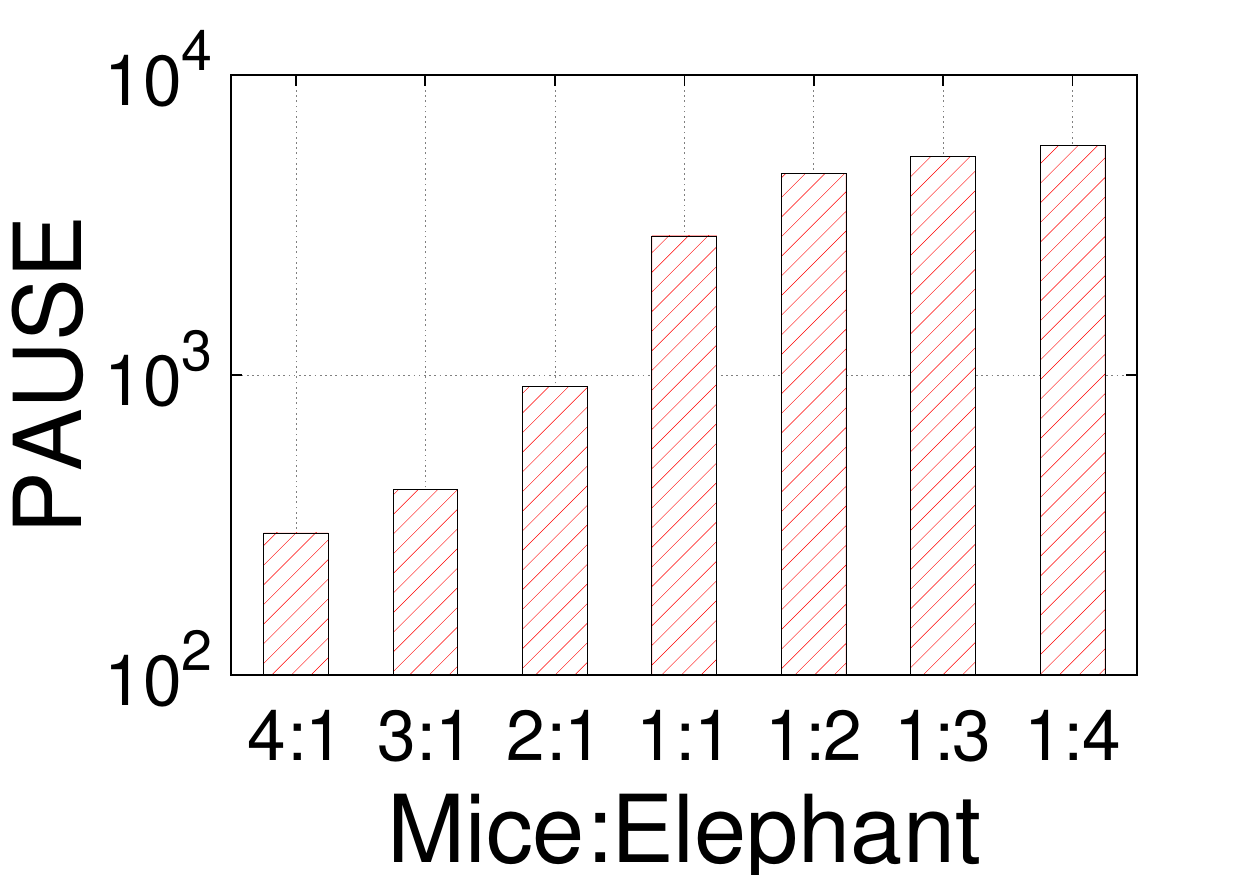}
	}
	\hfill
	\subfigure[Average feedback value for elephant]{
		\includegraphics[width=0.45\linewidth]{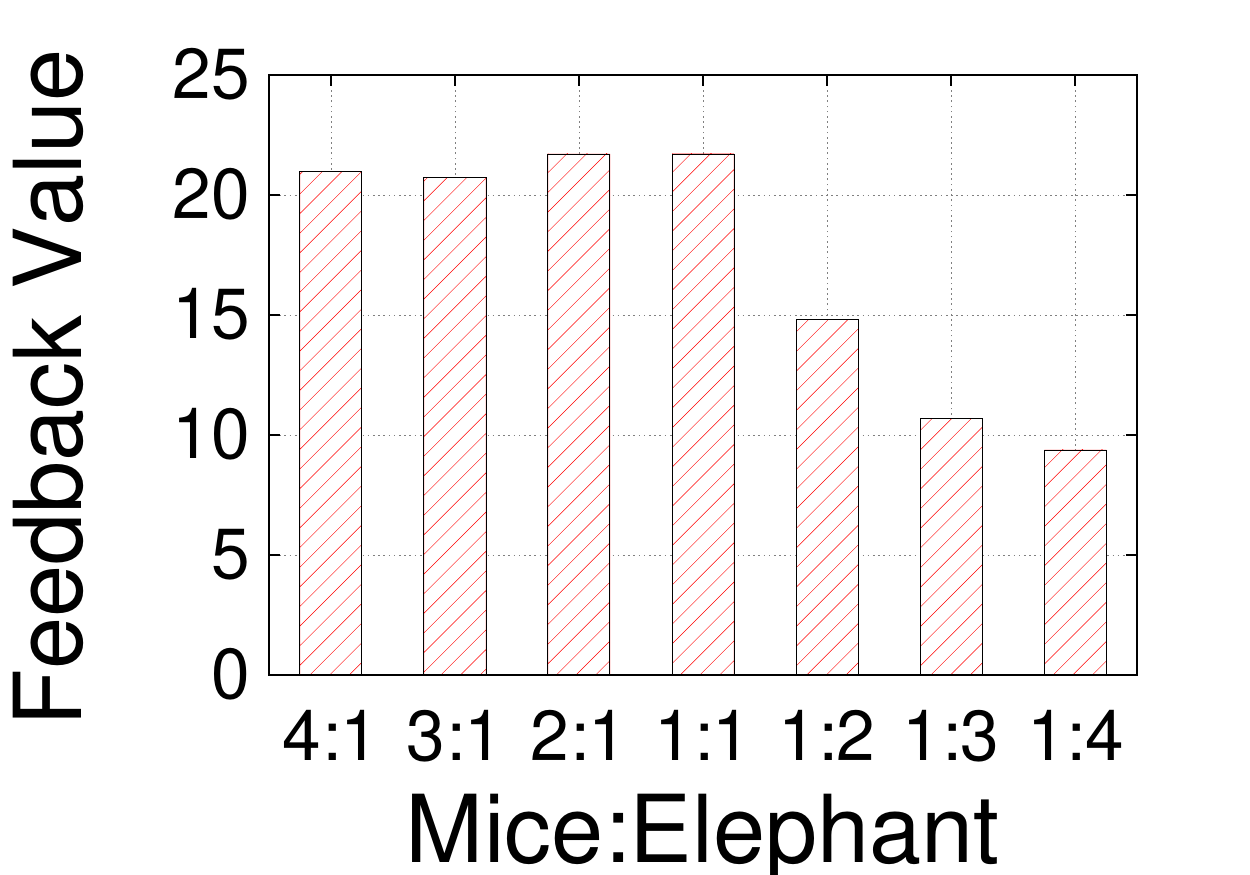}
	}
	\caption{Impacts of bandwidth allocation on the interaction between hop-by-hop and end-to-end flow control mechanisms. }
	\label{fig:ets-inter}
\end{figure}

Instead of the strict priority setting for mice and elephant, we configure fixed bandwidth allocation for these two traffic patterns.
Obviously shown in \emph{Fig.}\ref{fig:ets}, 
as the remained bandwidth for elephant increases, the latency of mice grows sharply while the bottleneck throughput varies within a scope of the small fluctuations and presents rising trend on the whole.
To verify our claims about the impacts of bandwidth allocation on the interaction between hop-by-hop and end-to-end flow control mechanisms, we record the total pause times and average feedback value for elephant (shown in \emph{Fig.}\ref{fig:ets-inter}).

Results in \emph{Fig.}\ref{fig:ets} and \emph{Fig.}\ref{fig:ets-inter} implies that the bandwidth allocation of mice and elephant impacts seriously on the latency of mice but only improves a little for the throughput of elephant.
But anyway, our isolation proposal improves the performance of mixture traffic compared with the original mixture configuration. 
We will use strict priority setting for mice and elephant in the rest evaluations.

\subsection{Upper layer end-to-end mechanisms}
Our proposal of isolation mainly consist of a hop-by-hop flow control mechanism for mice and a end-to-end flow control mechanism for elephant.
There are plentiful end-to-end flow control mechanisms appropriate for elephant. Except the L2 mechanism QCN, we can employ any other mechanism in our proposal of isolation, such as the upper layer mechanisms TCP and DCTCP.

\begin{figure}[tp]
	\centering
	\subfigure[Latency of mice]{
		\includegraphics[width=1\linewidth]{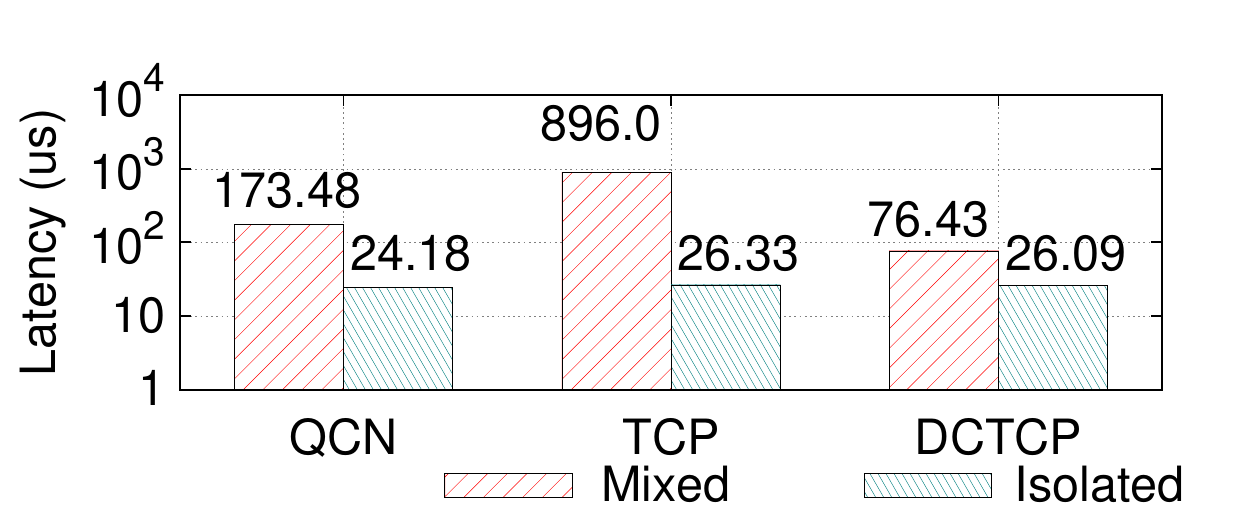}
	}
	\subfigure[Bottleneck throughput]{
		\includegraphics[width=1\linewidth]{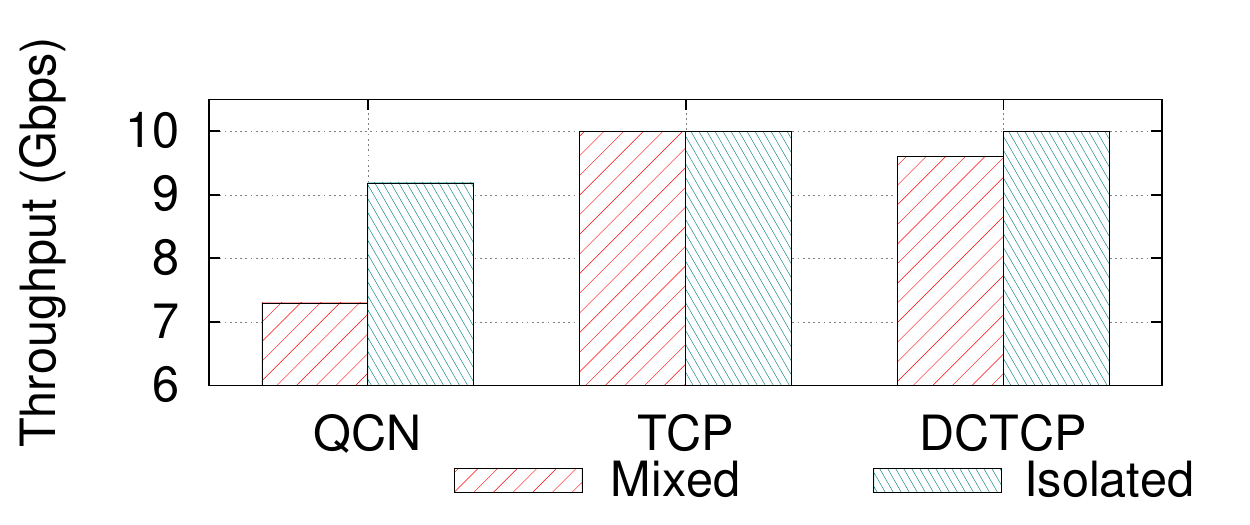}
	}
	\caption{Upper layer end-to-end mechanisms.}
	\label{fig:extend}
\end{figure}

Still in the many-to-one scenario, we evaluate the latency of mice and the bottleneck throughput using TCP and DCTCP together with PFC.
Comparison results of the traditional mixed configuration and our isolated configuration are shown in \emph{Fig.}\ref{fig:extend}.
For different end-to-end flow control mechanisms, there all exist performance impairments with mixture traffic, especially the bad latency for mice.   Our proposal of isolation can solve this problem that simultaneously improve the latency of mice and link utilization.

\subsection{Benchmark traffic}
In this subsection, we scale the leaf-spine topology to a large one with 18 ranks and 144 servers and evaluate our proposal the large-scale scenario.
Due to the bottleneck in data center is usually the tor switch \cite{vamanan2012deadline}, the mixture of short and long traffic in same rank will cause fairly different scenario compared with the mixture of cross-rank traffic. If the source and destination of all flows are in the same rank, the tor switch is worked as an inter-switching equipment. And in hence the bottleneck is the exchanging node in this scenario. On the other hand, if the source and destination of all flows are placed in different ranks, tor switch plays a role as the end point device in the entire network view. We design mainly two simulations which represent the conditions of above mentioned two scenarios respectively.
\subsubsection{Intra-rank evaluation}

In this evaluation, we compare the performance of isolation with mixed strategy with three widely used congestion control schemes (QCN, TCP and DCTCP). 8 servers are connected via simple tree topology, which can be seen as the structure of the single rank and the bottleneck is core switch. There is one server send query periodically (100us) to other servers. And remaining 7 servers respond short message (7.7KB) to the query, which constitute the short traffic pattern. Several inter-aggregation, long lived traffics are randomly generated at the same time. We evaluate the short flow completion time and bottleneck throughput. 

\begin{figure}[t]
	\centering
	\subfigure[Latency of mice]{
		\includegraphics[width=1\linewidth]{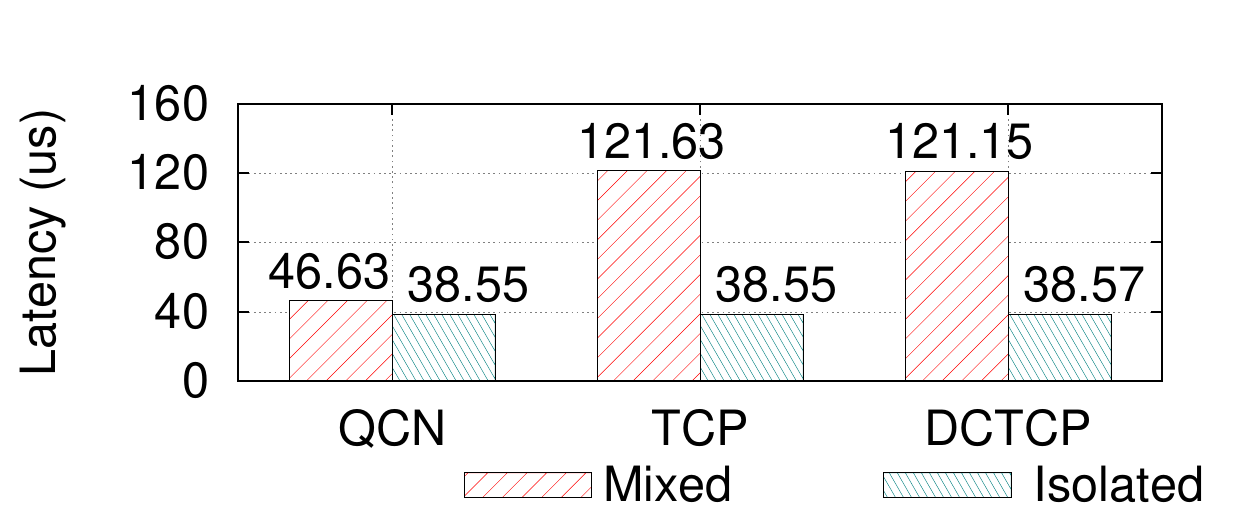}
		\label{fig:ev1-latency}
	}
	\subfigure[Bottleneck throughput]{
		\includegraphics[width=1\linewidth]{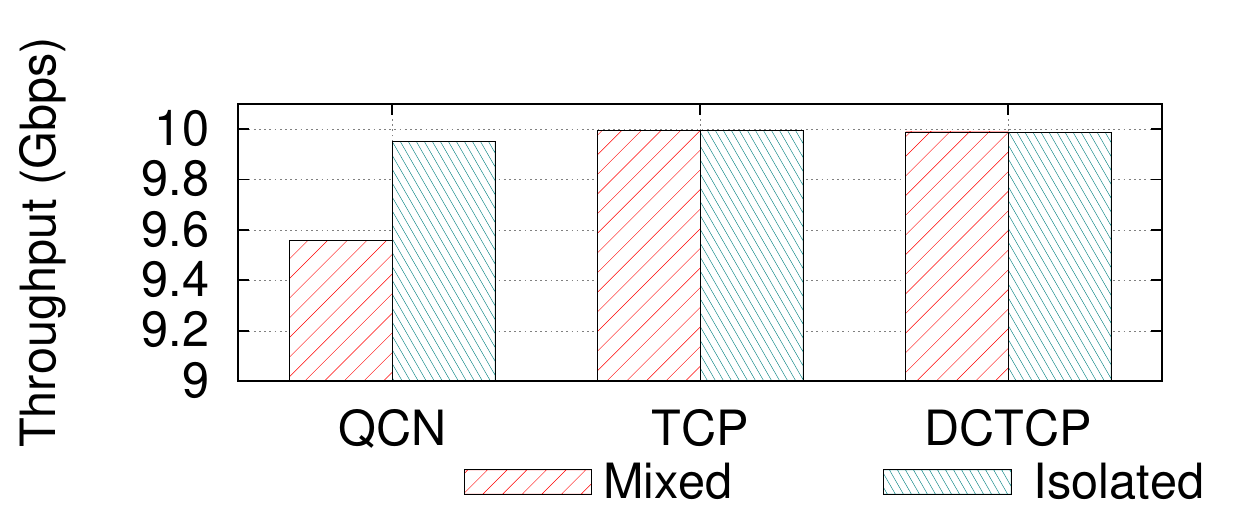}
		\label{fig:ev1-throughput}
	}
	\caption{Intra-rank evaluation.}
	\label{fig:ev1}
\end{figure}

\emph{Fig.}\ref{fig:ev1-latency} represents the completion time of mice flows in the isolation and mixture strategy respectively. The completion time of isolated mice flow is fairly low and steady under all three congestion control mechanism, which implies that the isolation is efficient to minimize the influence of long flows' queue length on the short messages' delay. In \emph{Fig.}\ref{fig:ev1-throughput}, isolation can always achieve full bandwidth utilization at the bottleneck. Notice that although in the mixture scenario QCN effectively decreases the mice completion time, it damages the bottleneck throughput greatly. By isolating mice and elephant, congestion control mechanism can take effect on the long-lived flows with limited influence from the short message burst. 

%
%

\subsubsection{Inter-rank evaluation}
In the inter-rank evaluation, we use the 144 nodes leaf-spine topology to model the real data center scenario. Two servers per rank send queries to 7 other servers in different ranks periodically (1ms) and those servers delivery 7.7KB length message as a response. The query-generating servers also maintain long lived traffic from inter-rank servers. There are some other long lived traffics, whose source and destination are randomly generated, in the network. The measured metrics are the same in the above evaluation. Similar to the results in last part, our proposal of isolation reduce the latency of mice to a same low value. The link utilization under the original mixed method and our proposal are almost 100\% for we set random traffic that number of concurrent flows is quite small (about 2) in this scenario.  
\begin{figure}[t]
	\centering
	\subfigure[Latency of mice]{
		\includegraphics[width=1\linewidth]{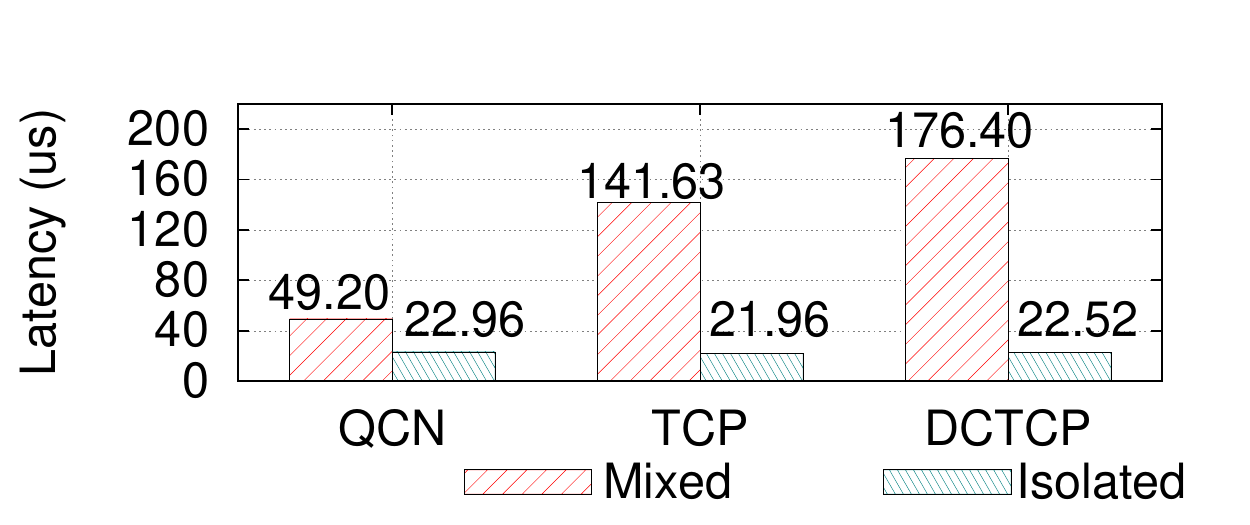}
		\label{fig:ev2-latency}
	}
	\subfigure[Bottleneck throughput]{
		\includegraphics[width=1\linewidth]{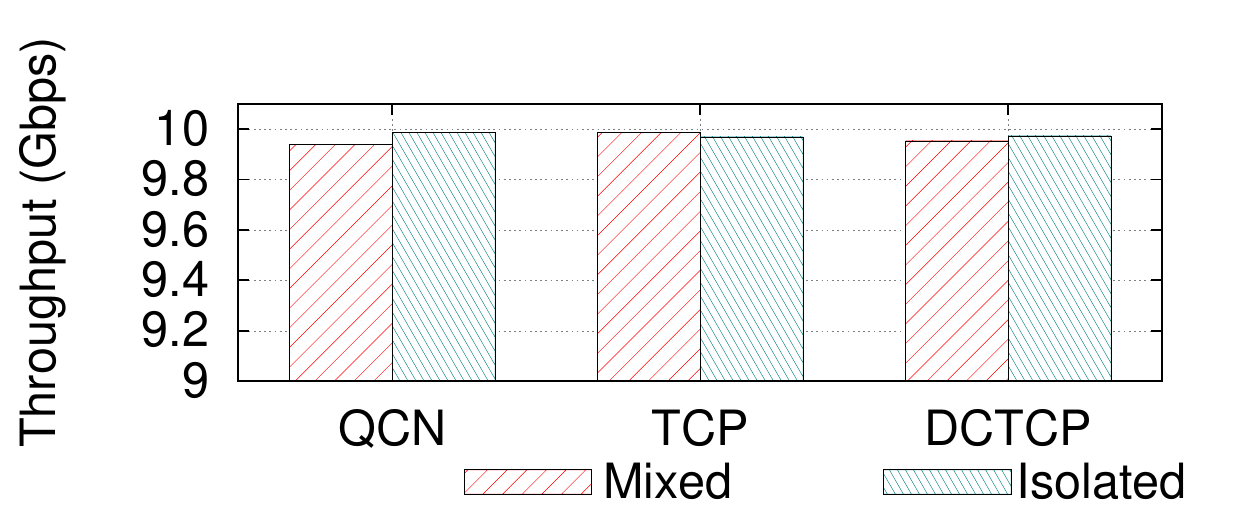}
		\label{fig:ev2-throughput}
	}
	\caption{Intra-rank evaluation.}
	\label{fig:ev2}
\end{figure}

\section{Related work}
\label{related}
Congestion management and control in data center networks is an active researcher area, we briefly summary the prior work into three rough categories: load balancing, transport protocol, in-network prioritization.

\textbf{Load balancing}: Hedera \cite{al2010hedera} and \cite{benson2011microte} use centralized traffic engineering to reroute flows to multipath, and both of them need handle traffic volatility properly.  CONGA \cite{alizadeh2014conga} splits TCP flows into flowlet, and allocate flowlets to paths based on network conditions. The interplay between CONGA and congestion control mechanism is an interesting open issue, and the swing decisions is a possible potential problem. Presto \cite{he2015presto} divides flows into equal-size ``flowcells'' at hosts, and sprays ``flowcells'' on the multipath in a round robin fashion irrespective of network congestion. The intensive packet reordering in asymmetric topologies is a heavy burden for high performance protocol stack.

\textbf{Transport protocol}: To minimizing flow completion time or meeting deadline for short flows, the transport protocol is customized for data center networks, including DCTCP \cite{alizadeh2011data}, D2TCP \cite{vamanan2012deadline} and TIMELY \cite{mittal2015timely} etc. Since employing the paradigm of end-to-end congestion control mechanism, they are naturally unfit for handling transient congestion caused by short flows, so it is unlikely to perfectly satisfy the performance requirements of latency-sensitive short flows. HULL \cite{perry2014fastpass} caps link utilization and employs both DCTCP and pacing mechanism to keep near baseline fabric latency. In high fan-in traffic pattern, the self-clocking mechanism in window-based flow control disables the ability of pacing in smoothing burst caused by concurrent flows, the ultra-low latency cannot be maintained. Both D3 \cite{wilson2011better} and Fastpass \cite{alizadeh2012less} adopt in-network bandwidth allocation schemes. D3 requires special switch and NIC hardware, and the performance bottleneck of Fastpass likely appears in the centralized arbiter in large-scale networks due to intensive signaling interaction and computing overheads.

\textbf{In-network prioritization}: PDQ \cite{hong2012finishing} pauses the contending flows and conducts Earliest Deadline First (EDF) scheduling to prioritize sluggish flows. The pausing readily creates a set of ``hotspot'' links that can lead to tree saturation. p-Fabric \cite{alizadeh2013pfabric} preemptively schedules flows using packet forwarding priorities in switches, but the extremely simple rate control readily makes a mismatch between injection traffic and network capacity, which resulting in packet dropping due to congestion and bandwidth wastage due to things done by halves. PIAS \cite{bai2015information} use priority queue to reduce flow completion time through Shortest Job First(SJF) scheduling. QJUMP \cite{grosvenor2015queues} allows the latency-sensitive high priority flows to ``jump'' over low priority flows, but limits high priority flows in relatively low rate, which imposes negative impact in latency-sensitive short message transmission.

In addition, both DeTail \cite{zats2012detail} and DCQCN \cite{zhu2015congestion} employ the hop-by-hop flow control in the link layer, i.e. PFC, but short and long flows are mixed in a queue. Isolating the flows with different size into the different queues is a common strategy, which is generally suggested by priority-based scheduling schemes, like PQD \cite{wilson2011better} and p-Fabric \cite{alizadeh2013pfabric} etc. In our work, isolating elephants and mice primarily intends to eliminate mutual damage in hop-by-hop flow control and end-to-end flow control, which is orthogonal and complementary to scheduling.

\section{Conclusion}
\label{conclusion}
Typical Traffic patterns in data centers are composed of numerous latency-sensitive ``mice'' flows and a few throughput-sensitive ``elephant'' flows, in which case we find the combination of QCN and PFC suffers from serious performance degradation. We conduct extensive simulations and in-depth analysis on all possible combinations of flow mechanisms and traffic patterns. We find that the root cause is the congestion signals of the two flow control mechanisms are polluted when mice and elephant interacts in one queue. In addition, through analysis we also find that PFC is enough for mice and QCN is enough for elephant.
Motivated by these insights, we propose a straightforward design of isolating mice and elephant in individual queues and respectively imposing hop-by-hop and end-to-end flow control mechanisms. Our proposal can be readily implemented without any changes to hardware and can be extended to other congestion control mechanisms. Extensive simulations validate the effectiveness of our proposal: the latency of mice and the link utilization simultaneously improve. 
Therefore,  we advocate to mediate the war between mice and elephants in data centers through isolating them into individual queue

\bibliographystyle{abbrv}
\bibliography{isolating}
\end{document}